# Non-Maxwellian viscoelastic stress relaxations in soft matter

Jake Song,[1,2] Niels Holten-Andersen[1] and Gareth H. McKinley[2]

[1]Department of Materials Science and Engineering and [2]Department of Mechanical Engineering
Massachusetts Institute of Technology, Cambridge, MA 02139, USA

**Viscoelastic stress relaxation is a basic characteristic of soft matter systems such as colloids, gels, and biological networks. Although the Maxwell model of linear viscoelasticity provides a classical description of stress relaxation, the Maxwell model is often not sufficient for capturing the complex relaxation dynamics of soft matter. In this Tutorial, we introduce and discuss the physics of non-Maxwellian linear stress relaxation as observed in soft materials, the ascribed origins of this effect in different systems, and appropriate models that can be used to capture this relaxation behavior. We provide a basic toolkit that can assist the understanding and modeling of the mechanical relaxation of soft materials for diverse applications.**

Soft matter systems are characterized by dynamic fluctuations and rearrangements within the microstructure, which play an important role in the function of a wide range of soft materials, such as cell-matrix interactions in biology,[1,2] or energy dissipation and self-healing in engineered soft materials[3,4]. These rearrangement events give rise to a time-dependent response in the mechanical properties, i.e., viscoelasticity.[5] The measurement of viscoelasticity can thus provide meaningful information into the dynamics of the soft material of interest. This is most commonly done at bulk scales using a dynamic mechanical analyzer[6] or a rheometer[5], though microscopic-scale measurements can also be made via microrheology[7] and scattering[8].

The dynamics of soft matter systems are non-trivial, and are characterized by a viscoelastic response that is often more complex than the Maxwell model of linear viscoelasticity (hence the term "complex fluids"). The Maxwell model is a canonical model introduced in elementary studies of linear viscoelasticity, which is characterized by a single characteristic relaxation time. However, measurements on real soft matter systems commonly exhibit a viscoelastic response that underscores a broad distribution of relaxation modes. These distributions of relaxation processes are generally interpreted as arising from a wide range of structural length-scales or relaxation mechanisms by which stress can relax, though the exact origins of these mechanisms are often not clear. Indeed, there is a vast literature on the microscopic origins of the broadly distributed relaxation modes observed in a wide range of soft matter systems, as well as on the modeling strategies to capture such viscoelastic responses. Navigating this literature to distill meaningful insights from, and appropriate models for, rheological measurements of different soft materials can therefore be a considerable challenge.

The aim of the Tutorial is to provide an overview of non-Maxwellian viscoelastic relaxation in soft matter, including the microscopic origins of the relaxation responses observed in different classes of soft materials, and for making an informed choice on suitable models to capture these relaxation responses. We first briefly recall the properties of the Maxwell model, and establish some of the basic physical scenarios in which a single Maxwell model succeeds in capturing the essential physics underlying the stress relaxation of soft matter (Section 1). We then illustrate deviations from the single Maxwell model in soft matter and summarize the putative origins of such effects which have been proposed across different systems (Section II). We next establish a basic mathematical language for describing viscoelastic relaxation in soft matter (Section III). We highlight common relaxation functions (Section IV) and corresponding mechanical models (Section V) that can be used to model non-Maxwellian viscoelasticity in soft matter, and lastly, outline basic statistical considerations when applying these models for experimental data (Section VI).

## I. Maxwell model of linear viscoelasticity

Exponential decays of the form $\phi = \phi_0 \exp(-t/\tau_c)$, where $\phi$ is an observable thermo-physical property and $\phi_0$ is the initial value at $t = 0$, are often used to model relaxation events arising from simple dynamical processes. For instance, exponential decays are exact solutions to describe the correlation of Brownian motion of particles, the correlation of dipoles in the rotational diffusion of a polar molecule (giving rise to the Debye dielectric relaxation[9,10]), and the kinetics of first-order reactions. The characteristic time-constant for the relaxation process is denoted $\tau_c$ in this review, though $\lambda$ is also often used in the complex fluids literature.[11]

In linear viscoelasticity, exponential decays in stress are exact solutions to step strain deformations applied to the Maxwell model.[12] This model is eponymously named after James C. Maxwell, who showed that a linear mechanical combination of a Hookean spring and dashpot in series give rise to a viscoelastic stress relaxation governed by a single time-scale, $\tau_c$. We recall the essential results of the model here, but defer a detailed discussion of the Maxwell model to the references.[5,11]

The Maxwell model (Fig. 1A) has a constitutive relation:

$$\sigma(t) + \frac{\eta}{G}\frac{d\sigma(t)}{dt} = \eta \frac{d\gamma(t)}{dt} \qquad (1)$$

where $\eta$ is the (linear) Newtonian viscosity of the dashpot, $G$ is the (linear) Hookean elasticity of the spring, and their ratio describes a characteristic relaxation time $\eta/G = \tau_c$. The Maxwell viscoelastic model is often used to model stress relaxation processes arising from step strain perturbations, $\gamma(t) = \gamma_0 \mathcal{H}(t)$ (where $\mathcal{H}(t)$ is the Heaviside step function), or small-amplitude oscillatory strain perturbations, $\gamma(t) = \gamma_0 \sin(\omega t)$. Solving the constitutive relation in the case of a step strain experiment, we find that the relaxation modulus $G(t)$ of a soft material system described by Eqn. 1 exhibits an exponential decay (Fig. 1B):

$$G(t) = G_0 \exp(-t/\tau_c) \qquad (2)$$

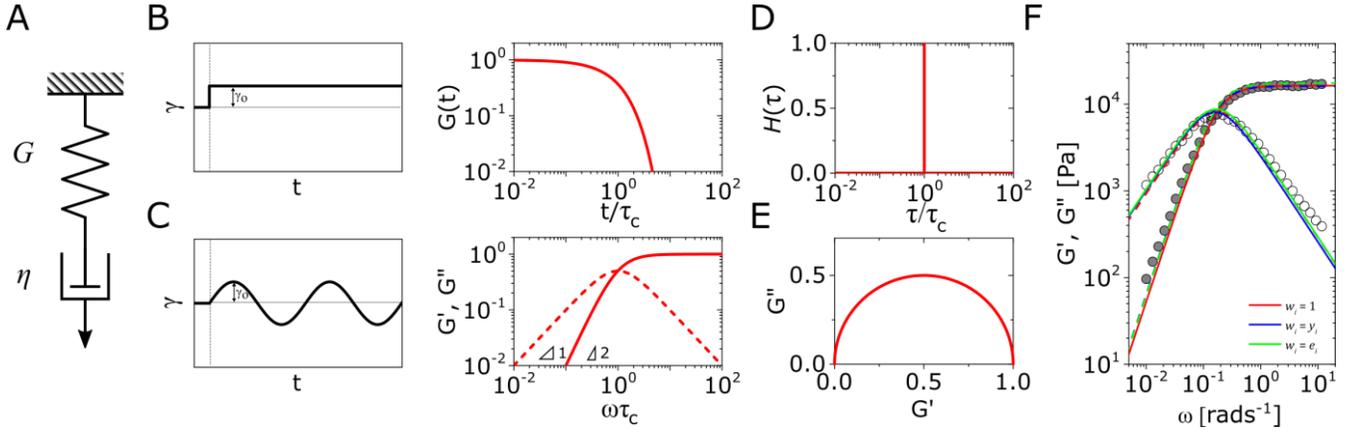

**Figure 1. The Maxwell model of linear viscoelasticity.** A) The Maxwell model consists of a spring with an elasticity of $G$, and a dashpot with viscosity of $\eta$, arranged in series. B,C) Linear viscoelastic responses of the Maxwell model to a step strain and oscillatory strain with $G_0 = 1$. D) The relaxation spectrum of $H(\tau)$ for the Maxwell model, represented by a delta function centered around $\tau = \tau_c$. E) A Cole-Cole plot of the loss modulus $G''(\omega)$ as a function of the storage modulus $G'(\omega)$. F) Transient polymer networks as an example of soft materials exhibiting near-Maxwellian viscoelasticity. This particular example shows the viscoelastic response of a metal-coordinating transient polymer network consisting of histidine-functionalized poly(ethylene glycol) with $Ni^{2+}$ ions.[13] Here we also demonstrate this fitting process using three different weights in the residual function – see Eqn. 26 in Section VI and discussion thereof – which leads to a statistical difference in the final fitting parameters.

where $G_0$ is the plateau modulus of the material. Solving the constitutive relation for the Maxwell model in the case of a linear oscillatory strain, the storage $G'(\omega)$ and loss $G''(\omega)$ components of the complex shear modulus $G^*(\omega)$ have the forms (Fig. 1C):

$$G^*(\omega) = G'(\omega) + iG''(\omega) = \frac{G_0}{1+i\omega\tau_c} \quad (3)$$

or by separating the real and imaginary components:

$$G'(\omega) = G_0 \frac{(\omega\tau_c)^2}{1+(\omega\tau_c)^2}$$
$$G''(\omega) = G_0 \frac{\omega\tau_c}{1+(\omega\tau_c)^2} \quad (4)$$

These materials functions exhibit a characteristic cross-over at a frequency of $\omega_c = 1/\tau_c$, with terminal power-law slopes at low frequency ($\omega \ll \omega_c$) of 2 and 1 for the storage and loss moduli, respectively. The Maxwell model response for oscillatory strain is thus dictated by the Deborah number arising from the oscillatory strain protocol, $De = \omega\tau_c$. The Deborah number helps characterize the importance of linear elastic effects in the system at different deformation rates, with $De \ll 1$ indicating a viscous liquid-dominated response, $De \gg 1$ indicating an elastic solid-dominated response, and $De = 1$ indicating the point of cross-over between the two response regimes.

In addition to the typical frequency sweep plot showing the storage and loss moduli as a function of frequency (Fig. 1C), there are other methods for representing $G'(\omega)$ and $G''(\omega)$ data which are used in the literature. For instance, the storage and loss moduli are also sometimes represented in a cross-plot, i.e., $G''(\omega)$ vs $G'(\omega)$ – in similar vein to the plotting methods used for real and imaginary values of the electrochemical impedances (the Nyquist plot) and dielectric constants (the Cole-Cole plot). This cross-plot of the real and imaginary response to oscillatory strain for a Maxwell material is a perfect semicircle (Fig. 1E); this provides a useful and demanding test of the accuracy of the Maxwell model when analyzing the dynamical response of soft materials. Another useful plotting strategy for oscillatory strain data is to plot the loss tangent $\tan\delta = G''(\omega)/G'(\omega)$ versus the magnitude of the complex shear modulus $|G^*(\omega)|$, which is often referred to as the van Gurp-Palmen plot.[14] The van Gurp-Palmen plot is essentially a simpler representation of the dynamic modulus data, which reveals the solid or liquid-like response of the system (since $\tan\delta = 1/\omega\tau_c = 1/De$) without needing to decompose the modulus into a storage and loss component as done in Fig. 1C.

Yet another useful representation for describing the viscoelasticity of soft matter is the continuous relaxation spectrum $H(\tau)$, which encodes the distribution of relaxation modes in a given system (more detail in Section III).[15,16] The continuous relaxation spectrum is encoded in the hereditary integral formulation of linear viscoelasticity (via the Boltzmann superposition principle),[17] and describes the relaxation modulus via:

$$G(t) = \int_{-\infty}^{\infty} H(\tau)\exp(-t/\tau)d\ln\tau \quad (5)$$

and the storage modulus $G'(\omega)$ and loss modulus $G''(\omega)$ via:

$$G'(\omega) = \int_{-\infty}^{\infty} H(\tau)\frac{(\omega\tau)^2}{1+(\omega\tau)^2}d\ln\tau$$
$$G''(\omega) = \int_{-\infty}^{\infty} H(\tau)\frac{\omega\tau}{1+(\omega\tau)^2}d\ln\tau \quad (6)$$

As the implication of the Maxwell model is the existence of a single characteristic relaxation time, $\tau_c = \eta/G$, the continuous relaxation spectrum of the Maxwell model is a delta function centered at $\tau_c$, and can be written in the compact form $H(\tau)/G_0 = \delta(\tau - \tau_c)$, where the integral over the delta function is equal to 1 such that $\int_{0-}^{0+} \delta(x)dx = 1$ (Fig. 1D).

Finally, in response to a step stress of amplitude $\sigma_0$, the Maxwell model exhibits an instantaneous increase in creep compliance $J(t) = \gamma(t)/\sigma_0$ (thus exhibiting a retardation time of zero), followed by a linear increase in creep compliance $J(t)$ with time and given by the relation:

$$J(t) = \frac{1}{G} + \frac{t}{\eta} \quad (7)$$

To keep the Tutorial concise and focused on viscoelastic *relaxation* phenomena, we omit extensive discussion of creep behavior (which measures the *retardation* response of the material). However, the topic of this review still applies broadly to creep responses, as linear viscoelastic properties such as $G(t)$, $G^*(\omega)$, and $J(t)$ are interconvertible via the Boltzmann superposition theorem.[11,15] It is also noted that creep tests can be quite a natural choice for measuring time-dependent mechanical properties in soft matter when using stress-controlled devices (i.e. most commercial rheometers). Using creep compliances can also be particularly advantageous in certain situations, for instance in analyzing entanglement plateaus in polymers[18] or analyzing microrheological data[19].

Though most viscoelastic stress relaxation responses of soft materials are more complex than a single exponential decay (Eqn. 2), there are a few exceptional cases where essentially Maxwellian behavior is observed. A prototypical example of materials exhibiting Maxwellian viscoelasticity are transient polymer networks, such as those consisting of star poly(ethylene glycol) (PEG) which are end-functionalized with diverse binding motifs such as metal-coordinating end-groups (Fig. 1F),[13,20-27] dynamic-covalent end-groups,[28,29] and ionic end-groups.[30] These materials are of interest due to the biocompatibility of the PEG as well as the shear-thinning behavior that arises from the disruption of the network that arises at high shear rates, which facilitates applications ranging from injectable hydrogels to cell culture[28,31,32]. Maxwellian behavior is also observed in other network materials as long as they are governed by first-order kinetics, such as DNA nanostars with controlled base-pair interactions,[33] telechelic polymers with hydrophobic domains such as the hydrophobic-ethoxylate urethane (HEUR) system,[34,35] and worm-like micelles, which are dimer networks where the interaction lifetimes define a single relaxation time between an entangled solid and a fluid.[36-38]

The relaxation dynamics of these Maxwellian networks can be most readily understood within the framework of transient network theory.[29,39-45] This theory describes the relaxation of polymer chains which are reversibly cross-linked by non-covalent bonds that serve as junction points of a rubbery network. In this picture, the polymer chains become affinely stretched when the network is rapidly deformed (e.g. via step strain); the accrued stress is relaxed when bond dissociation occurs, allowing the stretched polymer chains to re-associate with other polymer chains in a relaxed configuration. The relaxation process is thus governed by the bond dissociation time (which is often modeled by an Arrhenius-type expression), as well as by the density of elastically-active chains, the elastic force-extension response of the chains,[39,41] and cooperative effects arising from having several interaction sites on a single polymer chain.[46]

## II. Deviations from Maxwellian relaxation in soft matter

Soft materials commonly exhibit complex non-Maxwellian rheological behavior. We show a representative selection of such responses in Fig. 2, which illustrate deviations from Maxwell viscoelasticity in a diverse range of soft matter systems such as polysaccharide networks,[47] muscle tissues and food systems, polymer glasses,[48] mucin hydrogels,[49] cytoskeletal networks,[50] colloidal gels,[51] supramolecular polymers,[52] worm-like micelles,[36] and foams.[53]

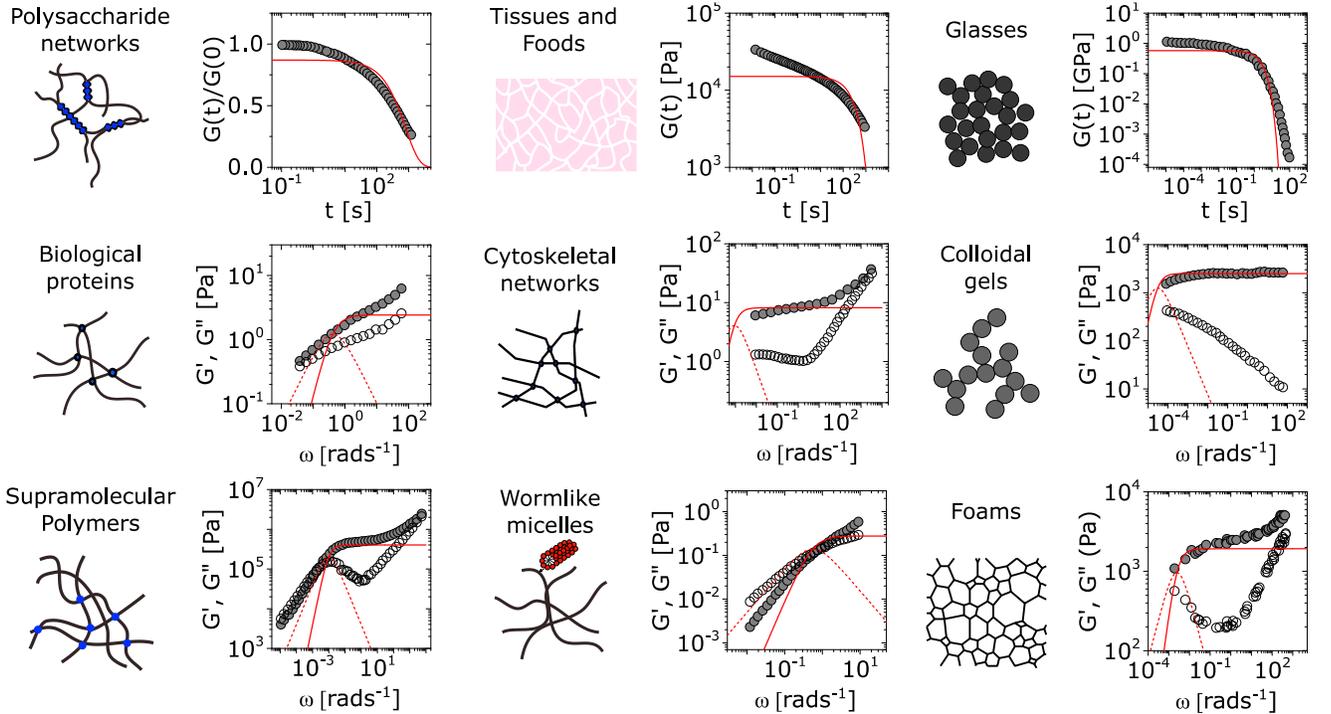

**Figure 2. A survey of deviations from Maxwellian viscoelasticity in soft matter.** The single-mode Maxwellian viscoelastic predictions are shown for comparison as solid lines for the relaxation modulus $G(t)$ and the storage modulus $G'(\omega)$, and dashed lines for the loss modulus $G''(\omega)$. All single-mode representations are obtained by non-linear fitting, and setting the weighting function $w_i = y_i$ (see Section VI). Shown here are the rheological behaviors of alginate reversibly cross-linked by $Ca^{2+}$,[47] muscle tissue of Yellowfin tuna, m-toluidine molecular glasses,[48], porcine gastric mucin,[49] actin-fascin networks,[50] silica colloidal gels,[51] supramolecular polymers bound by ureido-pyrimidinone moieties,[52] worm-like micelles of cetylpyridinium salicylate and sodium salicylate,[36] and shaving foam.[53] All data are digitized directly from original references, and are within the linear viscoelastic range as determined by the authors. The muscle tissue data are measured directly by performing shear rheology on a thin section of myotome of a Yellowfin tuna at room temperature (step strain $\gamma_0 = 0.5\%$).

Complex non-exponential relaxations are a hallmark behavior of a very wide range of soft materials, and numerous explanations of this phenomenon have been proposed. We briefly survey some of these explanations below.

### A. Non-first-order reaction kinetics

Deviations from Maxwellian viscoelastic relaxation can arise in telechelic networks when the bond interactions between polymers do not follow first-order reaction kinetics. A representative example of these kinds of interactions come from host-guest interactions, which are often structurally complex and can give rise to a complex dissociation pathway with many intermediate steps.[54] Indeed, studies have shown that multi-arm PEGs functionalized with cyclodextrin and cholesterol exhibit strong deviations from Maxwellian rheology,[55] in contrast to multi-arm PEGs functionalized with dynamic covalent linkages which exhibit Maxwellian linear viscoelastic responses.[28,32,56] The use of such complex interaction moieties provides an interesting, chemically-oriented approach to tuning the relaxation spectrum of the polymeric network.

### B. Chain dynamics

Deviations from single-mode Maxwell viscoelastic responses are also expected based on classical models of polymer dynamics. In the Rouse model, the relaxation dynamics are assumed to arise from the Brownian motion of beads connected by springs, wherein the terminal relaxation time depends on the polymer chain length. This leads to a stress relaxation response arising from a linear summation of individual (exponential) Rouse relaxation modes, which results in a power-law scaling of the relaxation modulus preceding a terminal relaxation, of the form:[57]

$$G(t) \sim G_0 (t/\tau_0)^{-1/2} \exp(-t/\tau_R) \qquad (8)$$

where $\tau_0$ is the shortest relaxation mode (of an individual Rouse segment) and $\tau_R$ is the longest relaxation mode of the entire chain. The Zimm model includes additional hydrodynamic interactions between individual segments of the bead-spring chain (and thus is more appropriate for dilute solutions of macromolecules where hydrodynamic effects are not screened). This changes the power-law scaling in Eq. 8 from $\sim t^{-1/2}$ to $\sim t^{-1/3\nu}$ where $\nu$ is the Flory scaling exponent that is related to the solvent quality.[57]

Non-Maxwellian stress relaxation becomes more prominent for polymer melts with long entangled chains, in which Rouse and Zimm modes are followed at longer times by entanglement effects.[58] This can result in the addition of stress relaxation mechanisms, such as reptation, contour-length fluctuations and constraint release.[59] More complex relaxation processes can also arise specifically due to entanglement effects in polymer systems of more complex topologies, such as entangled star polymers[60] and entangled ring polymers.[61]

Semiflexible polymers such as cytoskeletal polymers and extracellular matrix polymers have been shown to exhibit *unique* chain relaxation dynamics – distinct from Rouse and Zimm dynamics of polymers – which arises due to the significant thermal undulation of the semiflexible polymer backbone. Studies have shown that in such systems, the viscoelastic moduli exhibit a power-law scaling of $G'(\omega) \sim G''(\omega) \sim \omega^n$, in which $n = 3/4$ in the high frequency regime.[62-64]

### C. Sticky chain dynamics

Though chain dynamics such as Rouse dynamics are expected to occur even in swollen gels such as the Maxwellian transient networks introduced in Section I, Rouse relaxation modes occur at much shorter than those arising from the reversible interactions in these systems such that Rouse dynamics are often not measurable. When there is a sufficient concentration of interacting stickers per polymer chain in such transient networks, however, we may instead observe the manifestation of sticky chain dynamics.

A representative example of this is the sticky Rouse model,[42,46,65] in which the associations between polymer chains are understood to constrain the polymer chains and lead to an apparent Rouse-like dynamical regime at long times. This results in the observation of power-law relaxation of the form $G(t) \sim t^{-1/2}$ in associative polymers, most commonly in systems where the main chain has been functionalized by binding motifs such that multiple stickers exist per chain.[65-68] In similar vein, the concept of sticky dynamics have also been applied to reptation dynamics for studying associative networks with longer entangled chains.[69-71]

Sticky dynamics can also give rise to non-Maxwellian viscoelasticity in semiflexible polymer networks. Modeling studies have shown that dissociation events along the backbone of a semiflexible polymer generates transverse relaxation modes, in which successive relaxation modes become progressively slower as more dissociation events need to occur; this leads to a terminal power-law relaxation of $G(t) \sim t^{-1/2}$, which is commonly observed in semiflexible polymer networks.[72] Experimentally, this power-law region has also been shown to be characterized by stress fluctuations indicative of collective dynamics [73].

### D. Multiple relaxation modes

Non-Maxwellian viscoelastic responses are a natural outcome for systems which intrinsically have multiple relaxation modes, such as those with chain length polydispersity, sticker heterogeneity, or multiple relaxation mechanisms.

Polydispersity in the chain length can result in heterogeneous chain dynamics, and thus a broad distribution of relaxation modes. The observation of non-Maxwellian stress relaxations can be leveraged to gain insight into the polydispersity of polymer systems. For instance, for melt systems undergoing reptation, the observed distribution of relaxation modes (from $H(\tau)$) can be retraced to the molecular weight distribution of the system.[74]

For associative polymers, multiple relaxation modes can arise due to factors such as the sticker distribution on chains[75] and sticker clustering[76]. Stickers can also introduce spatial mismatches in the binding of two chains, resulting in formation of defects such as chain loops. This can generate energetic penalties in the chain and cause broadening in the viscoelastic relaxation curve.[77] These effects – especially in tandem with chain length polydispersity – can lead to complicated linear viscoelastic responses which can be challenging to ascribe to a single factor.

For composite materials such as polymer-particle systems, non-Maxwellian viscoelastic responses can arise due to multiple relaxation mechanisms that arise intrinsically from the composite microstructure. For instance, latex particles which are bridged by HEUR micelles exhibit a composite response of Rouse dynamics, bridging interactions, and large-scale cluster dynamics, the sum of which can lead to complex non-Maxwellian viscoelastic behavior.[78,79] Furthermore, each of these relaxation mechanisms can be governed by a distribution of relaxation modes. For the contribution arising from Rouse dynamics, such distribution can arise as a result of chain polydispersity. For the bridging dynamics, it is well understood that the number of bridging linkers can significantly change the terminal relaxation time of a pair of particles due to cooperativity.[80] We have recently shown that a Poisson distribution of linkers in associating particle systems can lead to a significantly non-Maxwellian viscoelastic response, which can be useful for understanding the dynamics of particle-polymer systems such as latex and HEUR, metal-coordinating nanoparticle hydrogels and coordination cage hydrogels.[81]. Finally, cluster dynamics can also be influenced by a distribution in cluster sizes, which we treat separately in the next Section IIE.

Lastly, a broad distribution in relaxation modes can arise in systems which are characterized by a distribution of activation energies. For instance, a Gaussian distribution in the activation energy of relaxation can facilitate a log-normal distribution in the stress relaxation time (see Section IVA). This approach has been taken to model the relaxation of a metal-coordinate polymer materials[82] (in which bond strengths are assumed to be distributed in a Gaussian manner) as well as glassy systems (in which relaxation is assumed to arise from local domains with a Gaussian distribution of activation energies).[83,84]

### E. Convolution of relaxation processes

A stretched exponential relaxation of the form

$$G(t) = G_0 \exp[(-t/\tau_c)^\beta] \quad (9)$$

is observed ubiquitously in soft materials such as glasses,[85-87] gels,[88] tissues[89], and surfactants.[36] Several studies have shown that this stretched exponential relaxation function is a direct outcome of relaxation processes which are convolved by a structural distribution of the relaxing unit. Mathematically, this can be expressed by an integral convolution of relaxation processes, such that:[90]

$$G(t) = G_0 \int_0^\infty \Phi(\tau) Q(t,\tau) d\tau \quad (10)$$

where $Q(t,\tau)$ is a function describing the relaxation, and $\Phi(\tau)$ is a probability function for a given relaxation time $\tau$ that is convolved with $Q(t,\tau)$. In the framework of soft matter relaxation, $Q(t,\tau)$ directly describes the exponential relaxation of the primary unit of a given system (e.g., reptation time of a chain or relaxation time of a cluster), and $\Phi(\tau)$ is related to a distribution in the topological features such as the length or size of the relaxing unit, which weights the size-dependent relaxation function $Q(t,\tau)$.

An example of this idea is the work of Douglas et al.,[91-93] which evaluates the relaxation function of polymeric systems that form clusters. In these works, it is assumed that the cluster relaxation time depends on the diffusion processes that scale with the size of the cluster. It is thus shown that the relaxation time scales as $\tau \sim L^2/D_0$ with $L$ being the characteristic length of the cluster and $D_0$ being the diffusivity of the cluster. Assuming that clusters take a Boltzmann (exponential) size distribution, one can derive the convolution function $\Phi(\tau) = \exp(-L) = \exp(-\tau^{1/2})$. A steepest descent approximation of Eqn. 10 with this convolution function results in the following derivation:

$$G(t) = G_0 \int_0^\infty \exp(-\tau^A) \exp(t/\tau) d\tau = G_0 \exp(-(t/\tau)^\beta) \quad (11)$$

where $A = 1/2$ and the stretching exponent $\beta = A/(A+1) = 1/3$.[90] Thus, cluster size polydispersity itself can lead to a stretched exponential of the form $G(t) \sim \exp(-(t/\tau)^{1/3})$, a form commonly seen in the stress relaxation of polymeric systems such as gels.[91-93]

A similar derivation has been proposed by Cates et al. for wormlike micelles undergoing reptation,[36] with the main difference being an additional assumption that the diffusivity $D_0 \sim 1/L$ arises due to the curvilinear diffusion of the chains along their confining tubes. The convolution function thus becomes $\Phi(\tau) = \exp(-L) = \exp(-\tau^{1/3})$, resulting in a final stretched exponential function of $G(t) \sim \exp(-(t/\tau)^{1/4})$.

Finally, a related approach is the work of Curro et al.,[94] who have shown that permanently cross-linked systems such as elastomers can also exhibit a power-law stress relaxation. This relaxation is ascribed to dangling chain ends in the system, such that $G(t) \sim \sum_{n=1}^{\infty} \sigma_n(t) P(n)$, where $n$ is the length of the chain branch, $\sigma_n(t) = n - l(t)$ describes the stress arising from parts of the dangling chains which have not relaxed ($l(t)$ is the lognormally-dependent relaxation rate of dangling chain ends as derived by de Gennes), and $P(n)$ is the probability of having a dangling chain of length $n$. The convolution of the probability of $n$ by the time-dependent stress arising from $n$ is shown to result in a power law relaxation $G(t) \sim t^{-(q/a)}$ where $q$ is the ratio of cross-link density to monomer density, and $a$ is a material constant related to the reptation time of the chain.[94] This idea of the disentanglement of dangling chain ends has also been utilized by Rubinstein et al. to explain the power-law stress relaxation in block copolymers, see reference.[95]

### F. Criticality and fractals

The rheological response of an associating soft material near critical points are often characterized by a power-law response in the frequency-dependent viscoelastic moduli such that $G'(\omega) \sim G''(\omega) \sim \omega^n$. This is observed in cross-linking polymers approaching gelation,[96] in soft colloidal systems approaching jamming,[97] and in fiber networks approaching critical connectivities.[98] Though these studies have canonically predicted $n = 1/2$ across the different systems, the actual exponent underlying the power-law relaxation has been shown to vary significantly depending on the specific nature of the system.

One important factor that appears to govern $n$ is the underlying fractal structure of the associated material. In branched polymeric materials, Muthukumar et al. has shown that $n$ can be related to the fractal dimension $d_f$ of the polymer network that forms at the percolation threshold.[99-101] Other works have shown that the fractal nature of the percolating system remains in the gel structure well beyond the percolation threshold, and that these structures result in remnant signatures in the loss modulus[102] and the relaxation spectrum of the resulting gel.[103,104] In marginal spring networks, the underlying fractal structure has also been associated with a value of $n$ that is substantially lower than $1/2$ at high frequencies,[105] though a direct relation between $d_f$ and $n$ is not yet clear in those systems.[106]

Another important factor is viscous coupling and hydrodynamics. In polymeric systems, hydrodynamic interactions lead to a different expression of $n$ as a function of $d_f$.[101] Through normal mode analysis of weak colloidal gels it has been shown that hydrodynamic interactions can drive a power-law relaxation response.[107] Finally, the viscous coupling of spring networks with solvents have also been shown to affect the value of $n$.[108] In all of these studies, it has been shown that hydrodynamic interactions result in an increase in $n$, accelerating the stress relaxation of soft materials.

### G. Colloidal gel dynamics

Several studies have also reported unique factors that give rise to the non-Maxwellian relaxation dynamics of colloidal gels. Cho et al. have shown that the density correlations within a single cluster of the gel follows a stretched exponential response[109,110]. A microrheological conversion[19] of this effect using the generalized Stokes-Einstein approach results in a Kelvin-Voigt like mechanical response, wherein the material acts like a solid at long times (low $\omega$) and a liquid at short times (high $\omega$). In such permanent systems there is another collective relaxation mode at longer times that can also manifest as a result of mutual constraints imposed by steric hindrance; this leads to subdiffusive dynamics, and thus a power-law decrease in $G'(\omega)$ and $G''(\omega)$.[111]

Zaccone et al., have shown that relating the cluster size distributions arising from attractive gelation of colloids to the continuous relaxation spectrum $H(\tau)$ can result in different relaxation spectra based on the fractal dimension of the associative colloidal gel (for instance, from gelation that proceeds via reaction-limited aggregation or diffusion-limited aggregation). Using this approach, a stretched exponential continuous relaxation spectrum $H(\tau)$ (see additional detail on $H(\tau)$ in Section IVA) is obtained for diffusion-limited aggregation, and a power-law $H(\tau)$ is obtained for reaction-limited aggregation and chemical gelation.[112]

### H. Non-linearity, non-affinity, and intermittency

A popular approach to understand non-Maxwellian relaxation in soft systems (particularly those that show aging dynamics) is through the soft glassy rheology (SGR) model[113,114], in which a power-law relaxation is obtained as a result of activated local yielding processes governed by an exponentially-distributed energy landscape. A "noise temperature" exponent can be extracted from the exponent of the power-law relaxation $x$ from $G'(\omega) \sim \omega^{x-1}$, which indicates how close the material is to a glass transition. The exponentially-distributed energy landscape is motivated by a trap model used for weakly aging systems[115]. This local yielding process can also be interpreted in terms of shear-transformation zones,[116-120] which has since become a popular method for interpreting relaxation dynamics of glassy and disordered systems.[121,122]

Studies have shown the importance of non-affine deformations on the power-law viscoelasticity of soft matter. This has been explored in particular detail in spring networks, which experience an increase in non-affine deformations near critical connectivity.[123] Studies have shown that such non-affine deformations can therefore directly affect the scaling response of the power law exponent governing $G'(\omega) \sim G''(\omega) \sim \omega^n$.[98,124] Recent works have also shown that such non-affine deformations can also arise in dense suspensions,[125] suggesting that the underlying physics governing spring networks and dense colloidal suspensions may be similar. In emulsions, it has also been shown that the power-law response in the loss modulus $G''$ can arise due to dissipation arising from non-affine motions[126]

Other studies have also demonstrated the importance of non-linear mechanical driving on the non-Maxwellian rheology of soft materials. and that non-linear mechanical stresses can significantly slow down the power-law relaxation observed in semiflexible polymers, with a corresponding decrease in the exponent $n$ in the relation $G(t) \sim t^{-n}$ from $n = 1/2$ as discussed previously in sticky semiflexible polymers, to a value as low as $n = 1/10$.[127] Our recent studies also support the picture of a non-linear mechanical process, where we show that the characteristic relaxation time $\tau_c$ of dynamically arrested soft materials such as gels are governed by internal stresses, and that mechanical driving leads to intermittent avalanches which result in a significant broadening in the continuous relaxation spectrum $H(\tau)$.[128] In this picture, the stretched-exponential-like stress relaxation of dynamically arrested solids are a manifestation of viscoplasticity arising from the marginal stability of these systems under imposed mechanical deformations.[129-131]

Studies have also shown that the scaling exponent governing the displacement trajectories of intermittent avalanches in materials that exhibit such marginal behavior (with power-law free energy landscapes) can be quantitatively linked to the scaling exponent of power-law stress relaxation of soft materials. In this framework, the superdiffusive exponent $\alpha$ obtained from the mean-square displacement relation $\langle \Delta r^2(\tau)\rangle \sim \tau^a$ can be related to the power-law exponent $n$ obtained from $G(t) \sim t^{-n}$. This requires the knowledge of the power spectrum of stress fluctuations $\langle \Delta \sigma^2(\tau)\rangle \sim \tau^\Delta$; derivations for arrested materials that exhibit spontaneous and random localized motion due to force dipoles (for instance, due to internal stresses[132,133]) result in $\Delta = 1$,[134] and a similar value is directly observed in coarsening foams[135] and cells.[136] The power-law exponent $n$ can be derived via a generalized Stokes-Einstein approach to obtain $n = (a - \Delta)/2$.[135]

## III. Introduction to modeling non-Maxwellian relaxations

The various proposed origins of non-Maxwellian relaxation processes in Section II invoke various functional forms of stress relaxation. We now introduce a basic mathematical toolkit to describe these non-exponential relaxation processes.

### A. The relaxation spectrum $H(\tau)$

The relaxation time spectrum $H(\tau)$ reveals the underlying relaxation modes governing the stress relaxation response. Each individual relaxation mode is assumed to be exponential, though strictly speaking they can be modified to take other forms, for instance a compressed exponential form.[128] It is also possible to interconvert $H(\tau)$ into a retardation spectrum $L(\tau)$,[18] and either of these spectra can be used to evaluate $G(t)$ or any other linear viscoelastic functions through the Boltzmann superposition integral (see Eqn. 5 and 6).[137-140] Thus, $H(\tau)$ encodes the distribution of relaxation modes which drive the linear viscoelastic behavior of a system. A Maxwellian system exhibits a single characteristic relaxation time $\tau_c$ corresponding to a $\delta$ function representation of $H(\tau)$ (Section I), while a non-Maxwellian system will exhibit a broader distribution of $H(\tau)$, thus allowing one to diagnose the statistical origins of non-Maxwellian responses in the rheological responses of the soft material in question.

Deriving the expression for $G(t)$ from a known distribution of $H(\tau)$ is fairly simple, and involves a simple numerical integration of the known functional form of $H(\tau)$. The reverse is not true, however. Obtaining $H(\tau)$ from $G(t)$ or $G^*(\omega)$ (which are inherently of limited temporal or frequency ranges) is challenging as it requires inverse Laplace transformations. This procedure is ill-posed, meaning that small variations in $G(t)$ can cause large variations in $H(\tau)$. This has mandated the use of approximation-based methods to obtain estimates of $H(\tau)$ from experimental data. There are various strategies to perform this operation,[138,140-142] but the use of Tikhonov regularizations is arguably the most established. This was first demonstrated in problems of viscoelasticity by Honerkamp et al[143], but the regularization method is commonly used in other disciplines as well.[144] The regularization method works by finding the optimal solution for $H(\tau)$ that minimizes the square error of the solution as well as a measure of the roughness of the resulting solution; for the sake of brevity we defer to references for more detailed instruction on the application of the method.[143-145] Regularization-based methods for obtaining $H(\tau)$ are now readily available for users via commercial rheometer software and independent codes[145].

The relaxation spectrum can be used in the form of discrete modes – in which the relaxation curve is deconstructed into a "line spectrum" of Maxwell modes[146] – or in the form of a continuous curve.[137,138] Using discrete relaxation modes can be advantageous when the viscoelastic relaxation can be associated directly with well-known underlying relaxation processes, for instance in telechelic metal-coordination-based polymer networks with multiple metal-ligand complexes.[13] Otherwise, discrete modes can lead to an unnecessarily large number of fitting parameters, as we show in the next section. A continuous parameterized curve for $H(\tau)$ is a more compact method for studying the underlying relaxation processes arising from a stress relaxation curve, though directly estimating the distribution *a priori* from experimental data can be challenging due to the errors incurred from the inverse Laplace transformation process.[143,147] Alternatively, fitting the experimental data to common functional forms utilized in rheology (see Sections IV and V) and obtaining their analytical solutions of $H(\tau)$ can sometimes provide a clearer description of the relaxation dynamics that underlie an observed non-Maxwellian viscoelastic response in soft materials.

### B. Discrete relaxation spectrum and the generalized Maxwell model

The most straightforward method to model non-Maxwellian relaxation is to assume the presence of discrete Maxwell modes in the relaxation function. A mechanical arrangement in which $i$ Maxwell elements are arranged in parallel gives rise to a Prony series of the Maxwell relaxation equation in Eqn. 2 such that:

$$G(t) = \sum_{i=1}^{N} G_i \exp(-t/\tau_i) \quad (12)$$

and

$$H(\tau) = \sum_{i=1}^{N} G_i \delta(\tau - \tau_i) \quad (13)$$

As discussed previously, this modeling strategy is most useful when there is a known number of relaxation modes. Otherwise, statistical approximations must be made to determine the number of relaxation modes in the system.[148]

A useful heuristic approach is to assume the presence of a relaxation mode per decade of time.[149] We demonstrate the application of this from the relaxation data of a nanoparticle-crosslinked hydrogel which follows a stretched exponential relaxation behavior (Fig. 4A).[128] As shown in the data, taking $N = 5$ Maxwell modes results in an excellent fit to the data, but the utility of this method is poor as we now have 10 fitting parameters with no discernable physical meaning. One can employ more sophisticated methods to determine a "parsimonious" relaxation spectrum[150] with the minimal number of modes, for instance by adding a penalty factor for overfitting via a Bayesian information criteria[151]. Of course, a pure statistical approach to model selection will be agnostic to the plausibility of the implied physics of the system and therefore must be used with care. There is ongoing research to incorporate such physical insights into statistical models (for instance, increasing the statistical likelihood of models in which the parameters can be restricted to a narrower range of values based on physical insight).[148]

### C. Continuous relaxation spectrum

A continuous relaxation spectrum provides a quantitative basis for interpreting the underlying physics of the relaxation phenomenon. Obtaining the continuous relaxation spectrum can be challenging, however, as the regularization process to convert rheological data to $H(\tau)$ as described above can lead to noisy results unless the collected data can cover a wide range of times or frequencies. Thus, a useful strategy is to either fit the rheological data to analytical solutions of microscopic models with well-known $H(\tau)$. For some functional forms of $H(\tau)$ that do not have a simple analytical function for $G(t)$ or $G^*(\omega)$, a direct numerical integration via Eqn. 5 and 6 can provide the relaxation modulus or the real and imaginary components of the complex shear modulus. A demonstration of this can be seen in the reference,[82] where a log-normal distribution in sticker strength is assumed to model the rheological response of an associative polymer system.

### D. Mittag-Leffler functions

We lastly introduce here the Mittag-Leffler (ML) functions, which are a family of generalizing functions that include the exponential function. This function occurs frequently as a solution to fractional order differential and integral equations in areas such as diffusive transport, chemical kinetics and viscoelasticity (see Section V for detail on fractional models of viscoelasticity).[10,152-158 159,160] The ML function of the variable $z$ can have up to three parameters $\{a, b, c\}$, and takes the form:[161,162]

$$E_{a,b}^{c}(z) = \frac{1}{\Gamma(c)} \sum_{n=0}^{\infty} \left[ \frac{z^n}{\Gamma(an+b)} \frac{\Gamma(c+n)}{n!} \right] \quad (14)$$

We obtain the two-parameter ML function when $c = 1$, the one-parameter ML function when $b = c = 1$, and the pure exponential function when $a = b = c = 1$. All of these variants of the ML function feature commonly in relaxation phenomena. The one parameter ML function represents an exact solution to anomalous diffusion,[163] the dielectric relaxation of the Cole-Cole model,[157,164] and the stress relaxation response of the fractional Maxwell gel model[156,158]. The two-parameter ML function features in the analytical solution for the stress relaxation of fractional Maxwell liquid and fractional Maxwell gel models[165,166]. The three-parameter ML function features in solutions of the relaxation dynamics of Cole-Davidson and Havriliak-Negami dielectric models[157,164]. Detailed mathematical descriptions of different Mittag-Leffler functions which are relevant for modeling complex relaxations are discussed in references.[152-154,157,162,164]

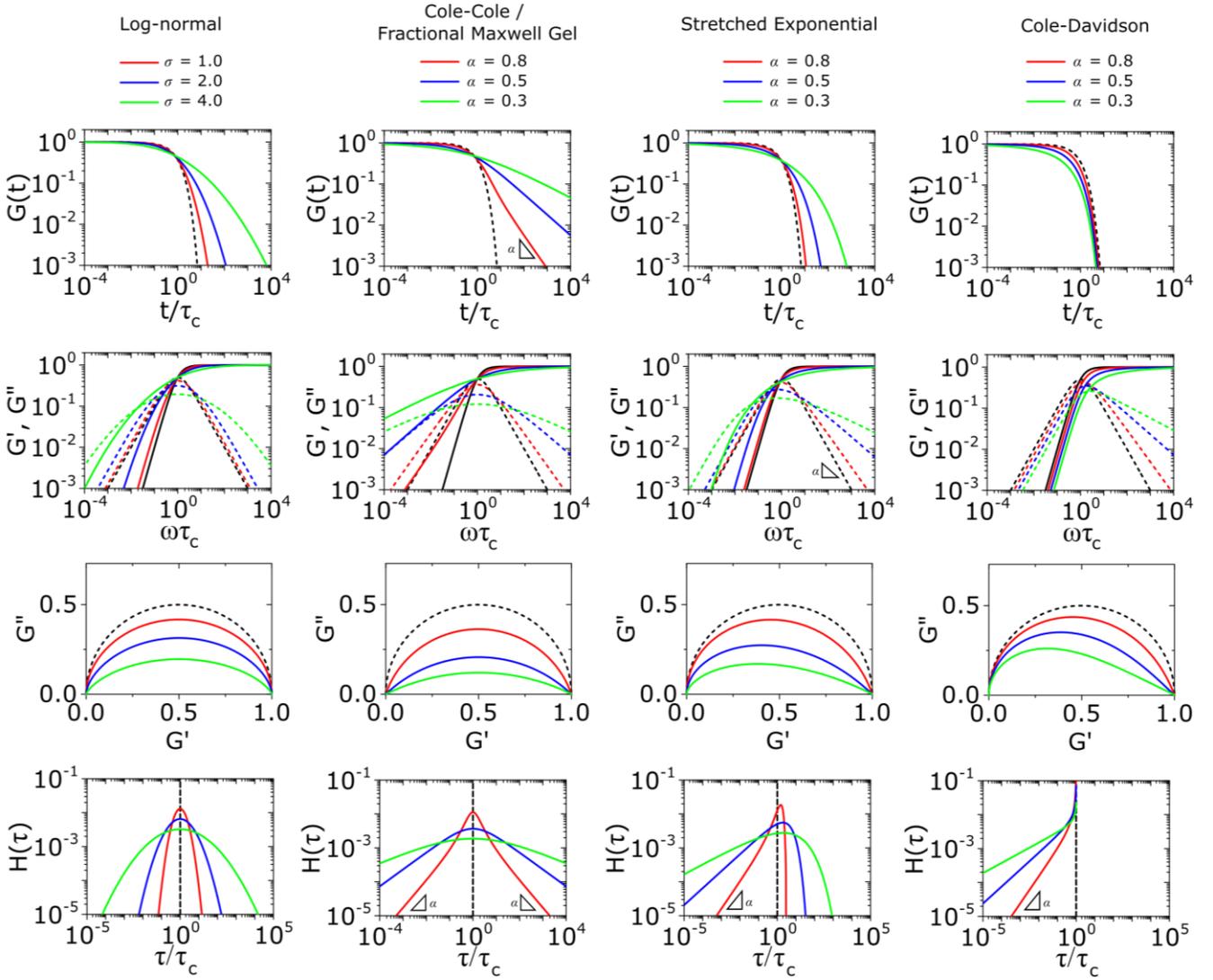

**Figure 3. The viscoelastic response of common relaxation functions for modeling non-Maxwellian viscoelastic responses introduced in the main text.** All moduli are normalized by the initial modulus $G_0$, and horizontally scaled by the characteristic relaxation time $\tau_c$.

## IV. Common relaxation functions

We apply the fundamentals laid out in Section III to highlight common functions used to model non-Maxwellian relaxation processes: the log-normal function, the Cole-Cole or fractional Maxwell gel function, the stretched exponential function, and the Cole-Davidson function. All functions here are valid under normalized conditions such that the initial modulus $G_0 = 1$.

### A. Log-normal function

A log-normal distribution of relaxation times has an analytical solution of the form:[167-170]

$$H(\tau) = \frac{G_0}{\sigma\sqrt{\pi}} \exp\left[-\left(\frac{\ln(\tau/\tau_c)}{\sigma}\right)^2\right] \quad (15)$$

where $\tau_c$ is the characteristic relaxation time (i.e., the most probable value) and $\sigma$ is the width of the distribution; analytical solutions for different $\sigma$ values are illustrated in Fig. 3. The log-normal relaxation time spectrum is useful in describing non-Maxwellian relaxation because it represents a direct solution of a Gaussian distribution of activation energies. This function may therefore be useful for describing spatially heterogeneous local activation energy of glasses[83,84,171], as well as modeling the polydisperse relaxation of reversible networks.[82] It may also be useful for modeling systems for which there is a Poisson distribution of relaxation modes, such as multivalent associative gels.[81]

The main challenge with using the log-normal function is the absence of easily implementable analytical solutions for the relaxation modulus or the storage and loss moduli. Therefore, $G(t)$ or $G^*(\omega)$ are usually evaluted by numerical integration of $H(\tau)$ in Eqn. 15 using Eqn. 5 and 6;[151,167,168,172-174] see Fig. 4C and 4D for demonstration.[82] We have also included a simple MATLAB protocol for performing this fitting procedures (see Resources).

### B. Cole-Cole and fractional Maxwell gel functions

The one-parameter version of the ML function with $z = -(t/\tau_c)^\alpha$ and $a = \alpha$ occurs frequently in the modeling of broadly distributed relaxation phenomena that can be described by fractional differential equations.[157,160] This ML function provides an analytical solution to the Cole-Cole model, which is a classical model used to describe deviations from the Debye dielectric relaxation model (the dielectric analogue to the Maxwell viscoelastic model), given by $\chi^*(\omega) = 1/(1 + (i\omega\tau_c)^\alpha)$.[175] An exact mathematical analogue to the Cole-Cole model exists in viscoelastic models in the form of the fractional

Maxwell gel model, which we introduce in greater detail later in the review.

The analytical description of the Cole-Cole or fractional Maxwell gel relaxation follows:[156,158]

$$G(t) = G_0 E_{\alpha,1}^1[-(t/\tau_c)^\alpha] \qquad (16)$$

where the one-parameter Mittag-Leffler function $E_\alpha[-(t/\tau_c)^\alpha]$ interpolates between a stretched exponential relaxation at short times ($t \ll \tau_c$) and a power-law decay in time at long times ($t \gg \tau_c$).[160] It can be shown using a Stieltjes transform (a complex generalization of the Laplace transform) that the relaxation function has a corresponding relaxation spectrum:[152,157,158]

$$H(\tau) = \frac{G_0}{\pi} \frac{(\tau/\tau_c)^{\alpha-1} \sin(\alpha\pi)}{(\tau/\tau_c)^{2\alpha} + 2(\tau/\tau_c)^\alpha \cos(\alpha\pi) + 1} \qquad (17)$$

which represents a symmetric distribution with power-law tails about the characteristic relaxation time.

The storage and loss moduli can also be derived analytically. The complex shear modulus $G^*(\omega)$ is described by the form:[156,158,159]

$$G^*(\omega) = G_0 \frac{(i\omega\tau_c)^\alpha}{1 + (i\omega\tau_c)^\alpha} \qquad (18)$$

By separating the real and imaginary part of $(i\omega\tau_c)^\alpha$ in Eqn. 18, it can be shown that the storage modulus $G'(\omega)$ and loss modulus $G''(\omega)$ take the form:[156]

$$\begin{aligned} G'(\omega) &= G_0 \frac{(\omega\tau_c)^{2\alpha} + (\omega\tau_c)^\alpha \cos(\alpha\pi/2)}{1 + (\omega\tau_c)^{2\alpha} + 2(\omega\tau_c)^\alpha \cos(\alpha\pi/2)} \\ G''(\omega) &= G_0 \frac{(\omega\tau_c)^\alpha \sin(\alpha\pi/2)}{1 + (\omega\tau_c)^{2\alpha} + 2(\omega\tau_c)^\alpha \cos(\alpha\pi/2)} \end{aligned} \qquad (19)$$

where, in the limit of $\alpha = 1$, Eqn. 19 reduces to the Maxwellian response of Eqn. 4.

We show numerical evaluations of the relaxation modulus, the dynamic moduli, and the relaxation spectrum for different values of $\alpha$ in Fig. 3. This relaxation function proves to be highly useful in capturing the viscoelastic relaxation of gel-like systems, as we discuss in Section VB and show in Fig. 6B.

### C. Stretched exponential function

The stretched exponential function is widely used to model relaxation in soft and disordered materials, is a direct solution to convolution-based integrals (Section IIE), and is most commonly known in the form of Eqn. 9. The Fourier transform of the stretched exponential does not have an analytical solution, and thus fitting the stretched exponential function to storage and loss moduli $G'$ and $G''$ also requires numerical integration of $H(\tau)$.[32] The relaxation spectrum can be expanded in a power series of the form:[176]

$$H(\tau) = -\frac{G_0}{\pi} \frac{\tau_c}{\tau} \left[ \sum_{k=0}^\infty \frac{(-1)^k}{k!} \sin(\pi\alpha k)\Gamma(\alpha k + 1)\left(\frac{\tau}{\tau_c}\right)^{\alpha k} \right] \qquad (20)$$

where $\tau_c$ is the relaxation time and $\alpha$ is the stretching exponent of the KWW function described in Eqn. 9. The result of the series expansion for various $\alpha$ values are illustrated in Fig. 3, as well as the normalized $G(t)$ and $G^*(\omega)$ responses which can be derived from the numerical integration of $H(\tau)$ using Eqn. 5 and 6. A demonstration of this numerical process can also be seen in the MATLAB code repository (see Resources). As can be seen in the graphical representation of $H(\tau)$, decreasing $\alpha$ from $\alpha = 1$ (which is the Maxwell limit) causes the spectrum to broaden, and deviate from a single discrete response to an asymmetric distribution. At $\tau < \tau_c$, the KWW function has a heavy tail, which cuts off upon approaching $\tau_c$. This functional form indicates that the KWW function is an excellent choice for capturing asymmetric relaxation in systems which show non-Maxwellian relaxation dynamics at short times below the characteristic relaxation time of the system, but which become progressively Maxwellian at long times. The stretched exponential function is commonly utilized for strongly arrested and glassy systems; we illustrate the application of this function for modeling the viscoelastic relaxation of polymer-nanoparticle gels in Fig. 4A.[128]

### D. Cole-Davidson function

A similar function to the stretched exponential function, the Cole-Davidson function is also sometimes used to capture asymmetric relaxation in materials which show a heavy-tailed distribution in relaxation modes at short times, and an abrupt single Maxwell relaxation mode at long times.[176,177] The function is less utilized than the stretched exponential function in modeling viscoelastic relaxation, but this function is commonly used in the modeling of dielectric relaxation.[157,176,178] We outline the analytical solutions below.

For mechanical relaxation, the Cole-Davidson function is most known through the functional form below:

$$G^*(\omega) = G_0\left(1 - \frac{1}{(1 + i\omega\tau)^\alpha}\right) \qquad (21)$$

which has an analytical solution for the storage and loss moduli of the form:[179]

$$\begin{aligned} G'(\omega) &= G_0\left(1 - \cos(\alpha\theta)\cos^\alpha(\theta)\right) \\ G''(\omega) &= G_0 \sin(\alpha\theta)\cos^\alpha(\theta) \end{aligned} \qquad (22)$$

where $\theta = \arctan(\omega\tau_c)$. The underlying relaxation spectrum $H(\tau)$ has the form:[157,176]

$$H(\tau) = \frac{G_0}{\pi} \sin(\pi\alpha)\left(\frac{\tau}{\tau_c - \tau}\right)^\alpha \qquad (23)$$

Lastly, the Cole-Davidson function also has an analytical solution for the relaxation modulus $G(t)$:[155,157]

$$G(t) = G_0\left(1 - (t/\tau_c)^\alpha\right) E_{1,\alpha+1}^\alpha(-t/\tau_c) \qquad (24)$$

where $E_{1,a+1}^a$ is the three parameter Mittag-Leffler function in Eqn. 14. The analytical solutions for the normalized $G(t)$, $G^*(\omega)$ and $H(\tau)$ for various $\alpha$ values of the Cole-Davidson function are illustrated in Fig. 3. As shown in the behavior of $H(\tau)$, this form of the Mittag-Leffler function allows modeling of relaxation behavior with abrupt transitions between a power-law response at short times, and an exponential decay at long times.

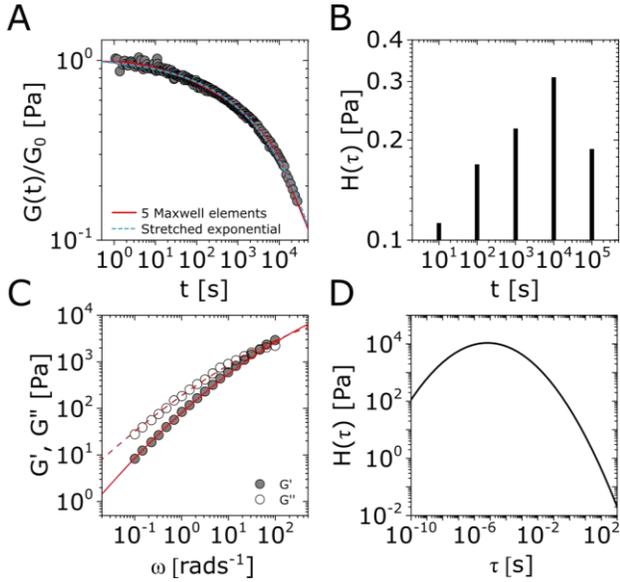

**Figure 4. Demonstration of common relaxation functions for modeling non-Maxwellian viscoelastic responses.** A) Stress relaxation modulus $G(t)$ of nitrocatechol-functionalized poly(ethylene glycol) networks reversibly cross-linked by $Fe_3O_4$ nanoparticles (time temperature superposition of data taken at temperatures $25^oC \leq T \leq 55^oC$)[128]. Shown alongside the data are the predictions of a generalized Maxwell model with $N = 5$ elements (one element per decade of time, as demonstrated in the classical reference.[149]), and a stretched exponential function. Both functions capture the relaxation behavior well, though the stretched exponential function (3) contains substantially less fitting parameters than the generalized Maxwell model (10). B) The obtained discrete relaxation spectrum $H(\tau)$ of the generalized Maxwell model used for fitting the data. The asymmetry in $H(\tau)$ is in agreement with the good fit of the data to the stretched exponential function which also has an asymmetric $H(\tau)$ (Fig. 3). C) Storage modulus $G'(\omega)$ (solid symbols) and loss modulus $G''(\omega)$ (open symbols) of spiropyran-functionalized polymer networks with transition metal cross-linking junctions.[82] A fitting procedure based on the log-normal distribution of relaxation modes $H(\tau)$ results in a good fit to the storage and loss modulus (solid and dashed lines, respectively). D) Corresponding log-normal $H(\tau)$ underlying the fitted results in C). All fitting parameters are shown in Table S1 in the Supplementary Information.

## V. Fractional mechanical models

We now introduce a family of mechanical models which are particularly useful for capturing the myriad of power-law relaxation responses in soft materials. This entails the use of a *spring-pot* mechanical element, which can interpolate between a spring (which has a constitutive equation $\sigma = G \frac{d^0\gamma}{dt^0} = G\gamma$) and a dashpot (which has a constitutive equation $\sigma = \eta \frac{d^1\gamma}{dt^1} = \eta\dot{\gamma}$). This is done through a fractional differentiation of strain with respect to time, such that:

$$\sigma = \mathbb{V}\frac{d^\alpha\gamma}{dt^\alpha} \quad (25)$$

where $0 \leq \alpha \leq 1$. $\mathbb{V}$ is a "quasi-property"[180] which interpolates between a spring-like response in $G$ and dashpot-like response in $\eta$, and has the units of $Pa \cdot s^\alpha$ (an excellent polar representation of the quasi-property in terms of $G$ and $\eta$ can be found in page 68 of[181]). Though the exact physical meaning of the spring-pot may appear nebulous, it has been shown that the fractional constitutive behavior of the spring-pot can be derived exactly with an infinite ladder arrangement of springs (with spring constants of $G_1, G_2, ... G_N$) and dashpots (with viscosities $\eta_1, \eta_2, ... \eta_N$) with $N \to \infty$ (Table 1).[10,16,91,166,182-184] In the ladder arrangement, the fractional exponent $\alpha$ is dictated by the scaling of $G$ and $\eta$ as a function of $N$, such that $G \sim \eta \sim N^{1-2a}$.[184] It has also been shown that a fractal network of spring and dashpots can also fulfill this kind of power-law relaxation (the ladder model can be interpreted as a variant of a fractal system).[91,184] This makes the spring-pot particularly appealing for describing relaxation processes driven by fractal structures, such as in critical gels. Other effects that result in power-law relaxations have also been linked to fractional models, such as Rouse dynamics.[185]

This mechanical model is also convenient as one can incorporate spring-pots into any conventional spring-dashpot mechanical configuration. The constitutive mechanical behavior then can be directly solved with fractional calculus operations (through either Riemann-Liouville or Caputo operators), the details of which we defer to the references.[10,158 156,159,165] The relaxation functions for many of the fractional models feature the Mittag-Leffler functions introduced in Section IIID, which represents a natural solution to fractional differential equations, and is also encountered in other kinetics problems ranging from anomalous diffusion[160,163] to infectious disease modeling.[186] Fractional mechanical models thus represent a natural choice for modeling the relaxation of soft materials exhibiting fractional kinetics.[187]

We introduce the more commonly used models below. All analytical results - the constitutive relation, responses to common rheological perturbations such as $G(t)$, $G'(\omega)$, $G''(\omega)$, and $J(t)$, as well as the relaxation spectrum $H(\tau)$ are listed in Table 1, with representative forms shown in Fig. 5. We also defer the readers to the references for more detail on the mathematical properties and applications of these models.[10,159,160,165,166,188] Finally, we provide a basic repository of MATLAB codes which demonstrates the application of these models in the fitting of viscoelastic relaxation data (Resources).

### A. Spring-pot

The spring-pot represents the basic building block of fractional mechanical models.[10,165,166] Individually, it can produce a power-law response in the relaxation modulus (Table 1, Fig. 5). The model is thus quite useful in the modeling of critical gels, in which both $G'(\omega)$ and $G''(\omega)$ share a common power-law.[96] It is noted that critical gels are often also described by a gel strength parameter $S$;[189] this term can be described by $S = \mathbb{V}/\Gamma(1 - \alpha)$ where $\Gamma(x)$ is the complete Gamma function. The spring-pot is a natural choice for modeling critical gels, due to the fractal nature of both the mechanical configuration of the spring-pot as well as the fractal microstructure of critical gels. A spring-pot indeed provides an excellent fit to a critical mucus network, as we show in Fig. 6A.[190]

### B. Fractional Maxwell gel

When a spring-pot is used in place of a dashpot in the Maxwell model, we obtain the fractional Maxwell gel model. The viscoelastic response of the fractional Maxwell gel model follows exactly the Mittag-Leffler function described in Section IVB and Fig. 3.[156,158] In this sense, the fractional Maxwell gel is an exact mechanical analog to the Cole-Cole model[191] which has been historically used in modeling anomalous dielectric relaxation in complex materials.[157,160,164,192]

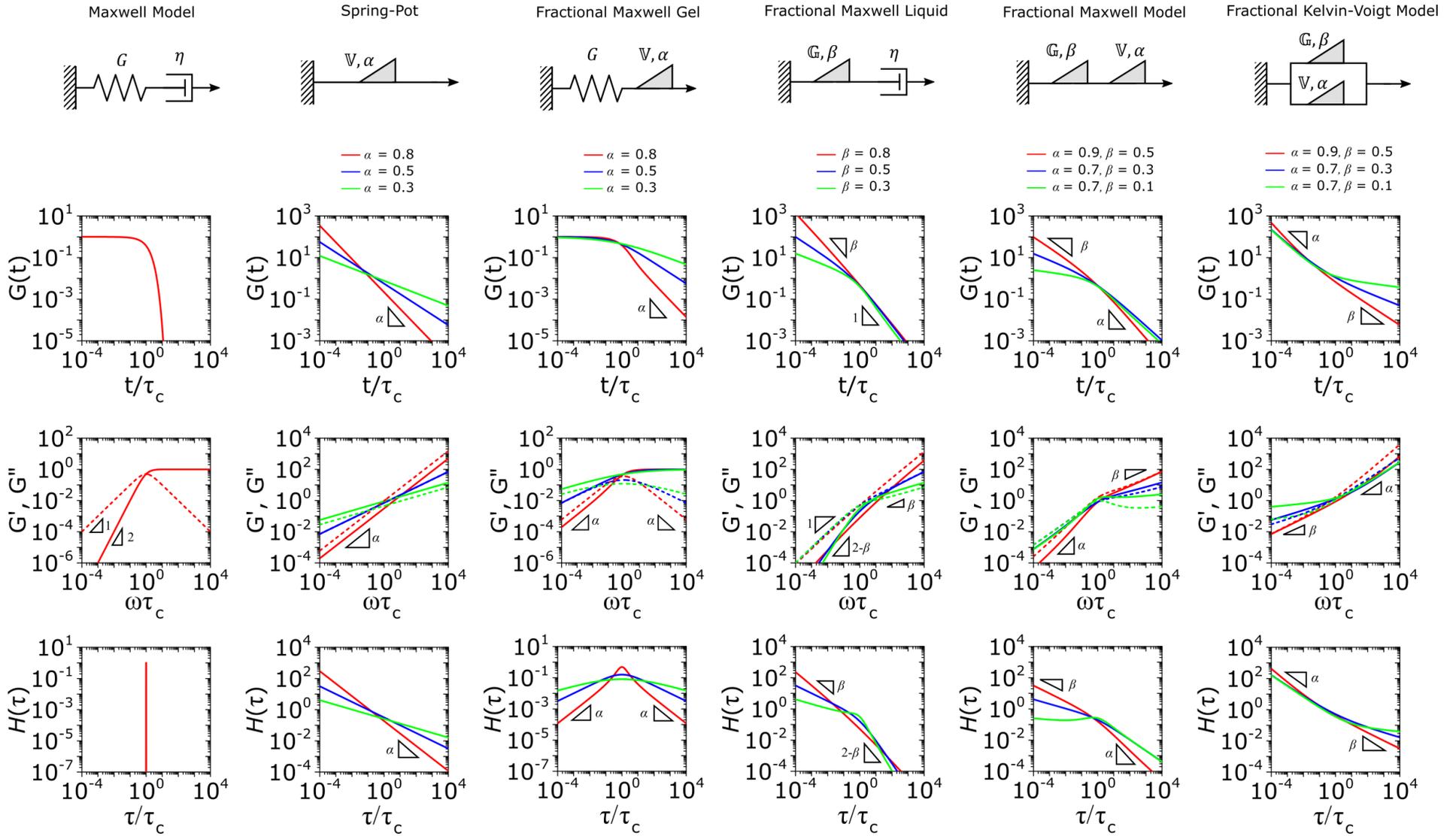

**Figure 5. Illustration of the viscoelastic response of the fractional mechanical models discussed in the main text.** All analytical functions corresponding to these graphic representations are listed in Table 1. The values of the characteristic modulus $G_c$ and characteristic relaxation time $\tau_c$ are set to unity for clarity.

.

The fractional Maxwell gel model enables the modeling a stress relaxation response of materials that exhibits a plateau modulus in a dynamic frequency sweep, and interpolates between a stretched exponential relaxation at short times and a power-law relaxation at long times.[160] It is thus useful for modeling gel-like materials[103,188,193] and we demonstrate the application of the fractional Maxwell gel model for fitting the $G'(\omega)$ and $G''(\omega)$ of a colloidal gel, see Fig. 6B.[51]

### C. Fractional Maxwell liquid

When a spring-pot is used in place of a spring in the Maxwell model, we obtain the fractional Maxwell liquid model. This model is useful for capturing the response of viscoelastic liquids which may exhibit a power-law relaxation mode preceding near-exponential relaxation (for instance, Rouse-like dynamics). We demonstrate this application on polyelectrolyte complexes in Fig. 6C.[194] It is worth noting, however, that the terminal relaxation in this model is not an exact exponential – because the relaxation follows a two-parameter Mittag-Leffler relaxation, the terminal slope of the storage modulus at low frequencies is $G'(\omega) \sim \omega^{2-\beta}$ where $\beta$ is the fractional exponent. Because the model contains a dashpot, the relaxation spectrum is integrable and yields a constant zero-shear viscosity.[195,196]

### D. Fractional Maxwell model

When both the spring and the dashpot are replaced by spring-pots, we obtain the fractional Maxwell model. This represents the most generalized linear viscoelastic model to capture mechanical responses to controlled strain (e.g., $G(t)$, $G'(\omega)$, $G''(\omega)$) as it combines the benefits of the fractional Maxwell gel and the fractional Maxwell liquid models; the fractional Maxwell model can easily be reduced into either of these models by setting one of the fractional exponents to 0 or 1, respectively. The stress relaxation response follows a two-parameter variant of the Mittag-Leffler function in Eqn. 14, and is useful for modeling viscoelastic materials which exhibit a smooth transition between two power-law regimes. This model is also particularly useful in modeling complex materials such as tissues and food composites[197,198]. We show a representative example of this in modeling the rheological response of the muscle tissues of Yellowfin tuna, which is more accurately modeled by the fractional Maxwell model than other functions such as the stretched exponential function, the log-normal function, and the generalized Maxwell model (Fig. 6D). In the limit of $\alpha = \beta \neq 1$, the model reduces to that of the single spring-pot.

### E. Fractional Kelvin-Voigt model, fractional Zener model, and beyond

We also provide a short introduction to the fractional analog of the Kelvin-Voigt model, which is the counterpart of the fractional Maxwell gel model for studying controlled stress responses. Indeed, just as the $G(t)$ response of the fractional Maxwell model followed a two-parameter Mittag-Leffler function, the creep compliance $J(t)$ follows a two-parameter Mittag-Leffler function. In terms of modeling strain-dependent functions such as $G'(\omega)$, and $G''(\omega)$, the fractional Kelvin-Voigt model may be useful in modeling rheological responses of soft materials exhibiting a high-frequency power-law. The fractional Kelvin Voigt can be further modified by adding an extra spring-pot to one of the series to create the fractional Zener model. For strain-dependent measurements (e.g., $G(t)$, $G'(\omega)$, $G''(\omega)$), the constitutive response of the fractional Zener model involves the simple addition of the response of the fractional Maxwell model and another spring-pot.[181] The fractional Zener model is quite useful in modeling viscoelastic responses of gels as well.[199] Overall, in the same way that traditional spring-dashpot configurations can be generalized into larger structures, fractional models can be generalized into larger structures, allowing flexible and customizable modeling of non-Maxwell viscoelastic relaxation in soft materials.

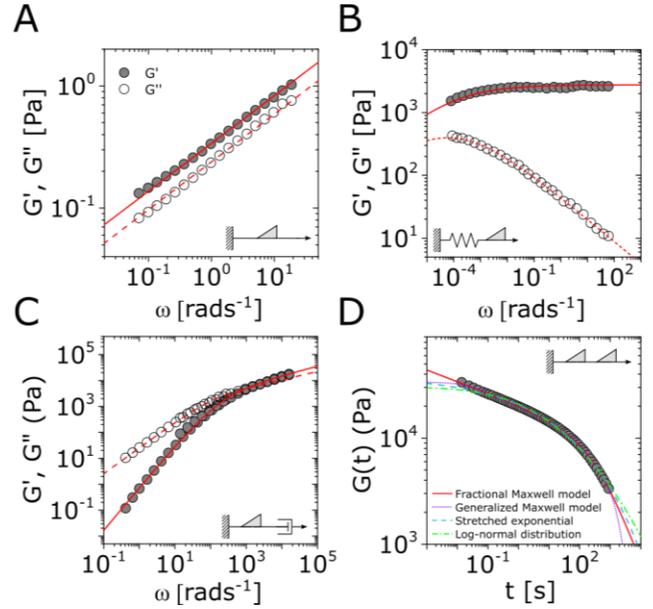

**Figure 6. Demonstration of the use of fractional mechanical models for modeling non-Maxwellian linear viscoelastic responses.** Viscoelastic responses of A) pig gastric mucin at pH = 4,[190] B) a colloidal silica gel[193], C) a poly(4-vinylpyridine) complex coacervate network (time-temperature superposition of data at $T = -7^oC$ and $T = 25^oC$)[194], and D) myotome (muscle) tissue of Yellowfin tuna (see Fig. 2 caption for experimental protocol). Excellent fits to these data are obtained using a spring-pot, a fractional Maxwell gel (FMG), a fractional Maxwell liquid (FML), and a fractional Maxwell model (FMM) (solid and dashed lines for $G'(\omega)$ and $G''(\omega)$ respectively, and solid line for $G(t)$). In D) we compare the fit to the fractional Maxwell model with fits to the stretched exponential function, the log-normal function, and the generalized Maxwell model; the fractional Maxwell model is deemed to be the most statistically likely as inferred from Bayesian information criteria (see attached MATLAB demo). All fitting parameters are shown in Table S1 in the Supplementary Information.

## VI. Statistical considerations in modeling rheological data

Using the many models introduced in the tutorial review to fit a given set of rheological data is process which is inherently governed by statistical methods. When the functional form of a model is known *a priori*, a common method to obtain the "right" parameters for the model that best describes a given set of data is to minimize the weighted residual sum of squares ($RSS_{w_i}$):

$$RSS_{w_i} = \sum_{i=1}^{n} \left( \frac{y_i - f(x_i)}{w_i} \right)^2 \qquad (26)$$

which computes the sum of the difference between data $y$ and model $f(x)$ which is then scaled by a weighting factor $w$. The choice of $w_i$ plays an important role in dictating the final parameters of a model, and we illustrate this idea using the Maxwellian viscoelastic data in Fig. 1F. Though the $RSS_{w_i}$ can be used without a weighting factor (i.e. $w_i = 1$), since rheological data is often logarithmic, the $RSS_{w_i}$ is conventionally rescaled by the magnitude of the data point such that $w_i = y_i$. However, statistical arguments suggest[200] that the most fundamental weighting factor is $w_i = e_i$, where $e_i$ can be obtained from the absolute uncertainty of measurements arising from a rheometer. Though the latter can be difficult to implement since $e_i$ needs to be measured directly on a rheometer through repeat measurements, Singh et al. have recently introduced an analytical estimation of $e_i$ based on rheometer specifications which simplifies the implementation of correct weighting factors in modeling rheological data.[200] In Fig. 1F, we illustrate the different outcomes of using $w_i = 1$, $w_i = y_i$, and $w_i = e_i$ to fit rheological data to the Maxwell model, which results in differences in the obtained fit values.

When the expected form of the relaxation process is not known *a priori*, one can infer the most likely model for a given set of rheological data using an appropriate statistical information criterion. Though the most straightforward method is to simply minimize the $RSS_{w_i}$, this method can be prone to overfitting.[148] This overfitting problem can be compensated by adding a term that penalizes large number of parameters. One such approach is the Bayesian information criterion (BIC):

$$BIC = -2\ln(\hat{L}) + n_p \ln(n_i) \quad (27)$$

where $\hat{L}$ is the maximum of the likelihood function, $n_p$ is the number of fitting parameters, and $n_i$ is the number of data points. For data arising from a Gaussian process with a known variance (for instance, $e_i$), this relation simplifies to:[201]

$$BIC = RSS_{w_i = e_i} + n_p \ln(n_i) \quad (28)$$

We can obtain the *BIC* values of the fractional Maxwell model, the generalized Maxwell model (with 5 elements), the stretched exponential function, and the log-normal function used to fit the stress relaxation data of Yellowfin tuna in Fig. 6D (see attached MATLAB demo and Fig. S1 in the supplementary information). We see that the fractional Maxwell model has the lowest *BIC* value of the four models, and thus represents the most statistically likely model. A similar operation on the small-amplitude oscillatory shear measurements on the metal-coordinating polymer network of Epstein et al. (from Fig. 4C) shows that the log-normal function is the most statistically likely model (Fig. S2).

These analyses also shows that the effect of the penalty term is relatively small compared to the likelihood term; the generalized Maxwell models ($n_p = 10$ for the tuna data and $n_p = 6$ for the metal-coordinating polymer network data) have a lower BIC value than the stretched exponential models ($n_p = 3$) in both cases, despite the stretched exponential exhibiting reasonably good fits to the data. This may be because the $e_i$ values used in this analysis – obtained from uncertainties arising from measured variables on the rheometer[200] – provides an under-estimation of the actual errors of the obtained data. More realistic estimations of errors must also consider uncertainties arising from experimental setup (for instance, sample underfilling and overfilling, step strain equilibration time) and from intrinsic heterogeneities in the samples, which would be challenging to quantify. Using larger weighting functions with Eqn. 27 will lower the magnitude of the likelihood term of the equation and increase the relevance of the penalty term.

## Conclusions

We have provided a summary on non-Maxwellian viscoelastic relaxation in soft materials, reviewing its diverse origins in different soft materials, and introducing mathematical models and statistical tools to model the observed relaxation responses. This tutorial review is aimed to guide appropriate model selections for capturing and decoding non-Maxwellian relaxation responses observed in a wide range of soft matter systems, and encourage future work on the physics underlying the complex relaxation phenomena in soft matter. A robust understanding of soft matter relaxation will have widespread implications in understanding and engineering natural and synthetic soft matter systems, respectively.

## Resources

We have uploaded a collection of MATLAB codes to demonstrate fitting small-amplitude oscillatory strain and step strain data to continuous relaxation spectra and fractional mechanical models to the File Exchange at https://www.mathworks.com/matlabcentral/fileexchange/111170-analysis-of-non-maxwellian-viscoelastic-data. Fitting results using these demonstrations can also be found in the supplementary information (Fig. S1 and S2).

## Acknowledgements

J.S. acknowledges financial support from the MIT Mathworks Fellowship. J.S. and G.H.M. would like to acknowledge R. H. Ewoldt (UIUC), P. K. Singh (Dow Chemical Company), J. D. Rathinaraj (MIT), and J. F. Douglas (NIST) for helpful discussions, and J-H. Cho (UCSB) for feedback on the manuscript.

## References

1. Chaudhuri, O., Cooper-White, J., Janmey, P. A., Mooney, D. J. & Shenoy, V. B. Effects of extracellular matrix viscoelasticity on cellular behaviour. *Nature* **584**, 535-546 (2020).
2. Hofer, M. & Lutolf, M. P. Engineering organoids. *Nature Reviews Materials* **6**, 402-420 (2021).
3. Blaiszik, B. J. *et al.* Self-healing polymers and composites. *Annual Review Of Materials Research* **40**, 179-211 (2010).
4. Campanella, A., Döhler, D. & Binder, W. H. Self-healing in supramolecular polymers. *Macromolecular Rapid Communications* **39**, 1700739 (2018).
5. Macosko, C. W. & Larson, R. G. *Rheology: principles, measurements, and applications*. (Wiley, 1994).
6. Menard, K. P. & Menard, N. R. *Dynamic mechanical analysis*. (CRC press, 2020).
7. Furst, E. M. & Squires, T. M. *Microrheology*. (Oxford University Press, 2017).
8. Scheffold, F. *et al.* New trends in optical microrheology of complex fluids and gels. *Trends in Colloid and Interface Science Xvi*, 141-146 (2004).
9. Debye, P. *Polare Molekeln*. (Hirzel, 1929).
10. Schiessel, H., Metzler, R., Blumen, A. & Nonnenmacher, T. Generalized viscoelastic models: their fractional equations with solutions. *Journal Of Physics A: Mathematical And General* **28**, 6567 (1995).
11. Bird, R. B., Armstrong, R. C. & Hassager, O. *Dynamics of Polymeric Liquids. Vol. 1: Fluid Mechanics*. 2nd edn, (Wiley, 1987).
12. Maxwell, J. C. IV. On the dynamical theory of gases. *Philosophical transactions of the Royal Society of London*, 49-88 (1867).
13. Grindy, S. C. *et al.* Control of hierarchical polymer mechanics with bioinspired metal-coordination dynamics. *Nature Materials* **14**, 1210 (2015).
14. Van Gurp, M. & Palmen, J. Time-temperature superposition for polymeric blends. *Rheol. Bull* **67**, 5-8 (1998).
15. Ferry, J. D. *Viscoelastic Properties of Polymers*. 80-98 (John Wiley & Sons, 1980).
16. Tschoegl, N. W. *The Phenomenological Theory of Linear Viscoelastic Behavior: An Introduction*. (Springer Science & Business Media, 2012).
17. Mours, M. & Winter, H. in *Experimental Methods In Polymer Science* 495-546 (Elsevier, 2000).

18. Plazek, D. J. & Echeverrıa, I. Don't cry for me Charlie Brown, or with compliance comes comprehension. *Journal of Rheology* **44**, 831-841 (2000).
19. Mason, T. G. Estimating the viscoelastic moduli of complex fluids using the generalized Stokes–Einstein equation. *Rheologica Acta* **39**, 371-378 (2000).
20. Grindy, S. C., Lenz, M. & Holten-Andersen, N. Engineering elasticity and relaxation time in metal-coordinate cross-linked hydrogels. *Macromolecules* **49**, 8306-8312 (2016).
21. Holten-Andersen, N. *et al.* Metal-coordination: using one of nature's tricks to control soft material mechanics. *Journal Of Materials Chemistry B* **2**, 2467-2472 (2014).
22. Holten-Andersen, N. *et al.* pH-induced metal-ligand cross-links inspired by mussel yield self-healing polymer networks with near-covalent elastic moduli. *Proceedings of the National Academy of Sciences* **108**, 2651-2655 (2011).
23. Parada, G. A. & Zhao, X. Ideal reversible polymer networks. *Soft Matter* **14**, 5186-5196 (2018).
24. Cazzell, S. A. & Holten-Andersen, N. Expanding the stoichiometric window for metal cross-linked gel assembly using competition. *Proceedings Of The National Academy Of Sciences* **116**, 21369-21374 (2019).
25. Rossow, T. & Seiffert, S. Supramolecular polymer gels with potential model-network structure. *Polymer Chemistry* **5**, 3018-3029 (2014).
26. Wang, R., Geven, M., Dijkstra, P. J., Martens, P. & Karperien, M. Hydrogels by supramolecular crosslinking of terpyridine end group functionalized 8-arm poly (ethylene glycol). *Soft Matter* **10**, 7328-7336 (2014).
27. Fullenkamp, D. E., He, L., Barrett, D. G., Burghardt, W. R. & Messersmith, P. B. Mussel-inspired histidine-based transient network metal coordination hydrogels. *Macromolecules* **46**, 1167-1174 (2013).
28. Yesilyurt, V. *et al.* Injectable self-healing glucose-responsive hydrogels with pH-regulated mechanical properties. *Advanced materials* **28**, 86-91 (2016).
29. Parada, G. A. & Zhao, X. Ideal reversible polymer networks. *Soft Matter* (2018).
30. Chassenieux, C. *et al.* Telechelic ionomers studied by light scattering and dynamic mechanical measurements. *Colloids And Surfaces A: Physicochemical And Engineering Aspects* **112**, 155-162 (1996).
31. Rosales, A. M. & Anseth, K. S. The design of reversible hydrogels to capture extracellular matrix dynamics. *Nature Reviews Materials* **1**, 1-15 (2016).
32. Tang, S. *et al.* Adaptable Fast Relaxing Boronate-Based Hydrogels for Probing Cell–Matrix Interactions. *Advanced Science* **5**, 1800638 (2018).
33. Conrad, N., Kennedy, T., Fygenson, D. K. & Saleh, O. A. Increasing valence pushes DNA nanostar networks to the isostatic point. *Proceedings Of The National Academy Of Sciences* **116**, 7238-7243 (2019).
34. Annable, T., Buscall, R., Ettelaie, R. & Whittlestone, D. The rheology of solutions of associating polymers: Comparison of experimental behavior with transient network theory. *Journal Of Rheology* **37**, 695-726 (1993).
35. Serero, Y. *et al.* Associating polymers: from "flowers" to transient networks. *Physical Review Letters* **81**, 5584 (1998).
36. Cates, M. & Candau, S. Statics and dynamics of worm-like surfactant micelles. *Journal Of Physics: Condensed Matter* **2**, 6869 (1990).
37. Cates, M. Reptation of living polymers: dynamics of entangled polymers in the presence of reversible chain-scission reactions. *Macromolecules* **20**, 2289-2296 (1987).
38. Rehage, H. & Hoffmann, H. Viscoelastic surfactant solutions: model systems for rheological research. *Molecular Physics* **74**, 933-973 (1991).
39. Tanaka, F. & Edwards, S. Viscoelastic properties of physically crosslinked networks: Part 1. Non-linear stationary viscoelasticity. *Journal Of Non-Newtonian Fluid Mechanics* **43**, 247-271 (1992).
40. Meng, F., Pritchard, R. H. & Terentjev, E. M. Stress relaxation, dynamics, and plasticity of transient polymer networks. *Macromolecules* **49**, 2843-2852 (2016).
41. Groot, R. D., Bot, A. & Agterof, W. G. Molecular theory of the yield behavior of a polymer gel: Application to gelatin. *The Journal Of Chemical Physics* **104**, 9220-9233 (1996).
42. Zhang, Z., Chen, Q. & Colby, R. H. Dynamics of associative polymers. *Soft Matter* **14**, 2961-2977 (2018).
43. Zhang, Z., Huang, C., Weiss, R. & Chen, Q. Association energy in strongly associative polymers. *Journal Of Rheology* **61**, 1199-1207 (2017).
44. Green, M. & Tobolsky, A. A new approach to the theory of relaxing polymeric media. *The Journal Of Chemical Physics* **14**, 80-92 (1946).
45. Bird, R. B., Curtiss, C. F., Armstrong, R. C. & Hassager, O. *Dynamics of polymeric liquids, volume 2: Kinetic theory*. (Wiley, 1987).
46. Rubinstein, M. & Semenov, A. N. Thermoreversible gelation in solutions of associating polymers. 2. Linear dynamics. *Macromolecules* **31**, 1386-1397 (1998).
47. Chaudhuri, O. *et al.* Hydrogels with tunable stress relaxation regulate stem cell fate and activity. *Nature Materials* **15**, 326 (2016).
48. Xu, B. & McKenna, G. B. Evaluation of the Dyre shoving model using dynamic data near the glass temperature. *The Journal of chemical physics* **134**, 124902 (2011).
49. Celli, J. P. *et al.* Helicobacter pylori moves through mucus by reducing mucin viscoelasticity. *Proceedings Of The National Academy Of Sciences* **106**, 14321-14326 (2009).
50. Lieleg, O. & Bausch, A. R. Cross-linker unbinding and self-similarity in bundled cytoskeletal networks. *Physical Review Letters* **99**, 158105 (2007).
51. Aime, S., Cipelletti, L. & Ramos, L. Power law viscoelasticity of a fractal colloidal gel. *Arxiv Preprint Arxiv:1802.03820* (2018).
52. Lewis, C. L., Stewart, K. & Anthamatten, M. The influence of hydrogen bonding side-groups on viscoelastic behavior of linear and network polymers. *Macromolecules* **47**, 729-740 (2014).
53. Gopal, A. & Durian, D. J. Relaxing in foam. *Physical Review Letters* **91**, 188303 (2003).
54. Velez-Vega, C. & Gilson, M. K. Force and Stress along Simulated Dissociation Pathways of Cucurbituril–Guest Systems. *Journal Of Chemical Theory And Computation* **8**, 966-976, doi:10.1021/ct2006902 (2012).
55. van de Manakker, F., Vermonden, T., el Morabit, N., van Nostrum, C. F. & Hennink, W. E. Rheological behavior of self-assembling PEG-β-cyclodextrin/PEG-cholesterol hydrogels. *Langmuir* **24**, 12559-12567 (2008).
56. Marco-Dufort, B., Iten, R. & Tibbitt, M. W. Linking molecular behavior to macroscopic properties in ideal dynamic covalent networks. *Journal of the American Chemical Society* **142**, 15371-15385 (2020).
57. Rubinstein, M. & Colby, R. H. *Polymer Physics*. Vol. 23 (Oxford University Press, 2003).
58. Doi, M. & Edwards, S. F. *The Theory of Polymer Dynamics*. Vol. 73 (Oxford University Press, 1988).
59. Likhtman, A. E. & McLeish, T. C. Quantitative theory for linear dynamics of linear entangled polymers. *Macromolecules* **35**, 6332-6343 (2002).
60. Milner, S. & McLeish, T. Parameter-free theory for stress relaxation in star polymer melts. *Macromolecules* **30**, 2159-2166 (1997).
61. Kapnistos, M. *et al.* Unexpected power-law stress relaxation of entangled ring polymers. *Nature Materials* **7**, 997-1002 (2008).
62. Gittes, F. & MacKintosh, F. Dynamic shear modulus of a semiflexible polymer network. *Physical Review E* **58**, R1241 (1998).
63. Morse, D. C. Viscoelasticity of tightly entangled solutions of semiflexible polymers. *Physical Review E* **58**, R1237 (1998).
64. Broedersz, C. P. & MacKintosh, F. C. Modeling semiflexible polymer networks. *Reviews Of Modern Physics* **86**, 995 (2014).
65. Chen, Q., Tudryn, G. J. & Colby, R. H. Ionomer dynamics and the sticky Rouse model. *Journal Of Rheology* **57**, 1441-1462 (2013).
66. Ahmadi, M., Hawke, L. G. D., Goldansaz, H. & van Ruymbeke, E. Dynamics of Entangled Linear Supramolecular Chains with Sticky Side Groups: Influence of Hindered Fluctuations. *Macromolecules* **48**, 7300-7310 (2015).
67. Tang, S., Wang, M. & Olsen, B. D. Anomalous self-diffusion and sticky Rouse dynamics in associative protein hydrogels. *Journal Of The American Chemical Society* **137**, 3946-3957 (2015).
68. Indei, T. & Takimoto, J. Linear viscoelastic properties of transient networks formed by associating polymers with multiple stickers. *The Journal Of Chemical Physics* **133**, 194902 (2010).
69. Rubinstein, M. & Semenov, A. N. Dynamics of entangled solutions of associating polymers. *Macromolecules* **34**, 1058-1068 (2001).
70. Nyrkova, I. & Semenov, A. Correlation effects in dynamics of living polymers. *EPL* **79**, 66007 (2007).
71. Schaefer, C., Laity, P. R., Holland, C. & McLeish, T. C. Silk Protein Solution: A Natural Example of Sticky Reptation. *Macromolecules* (2020).

72. Broedersz, C. P. *et al.* Cross-link-governed dynamics of biopolymer networks. *Physical Review Letters* **105**, 238101 (2010).
73. Müller, K. W. *et al.* Rheology of semiflexible bundle networks with transient linkers. *Physical Review Letters* **112**, 238102 (2014).
74. Pattamaprom, C. & Larson, R. G. Predicting the linear viscoelastic properties of monodisperse and polydisperse polystyrenes and polyethylenes. *Rheologica Acta* **40**, 516-532 (2001).
75. Shabbir, A. *et al.* Linear viscoelastic and dielectric relaxation response of unentangled UPy-based supramolecular networks. *Macromolecules* **49**, 3899-3910 (2016).
76. Ahmadi, M., Jangizehi, A., van Ruymbeke, E. & Seiffert, S. Deconvolution of the Effects of Binary Associations and Collective Assemblies on the Rheological Properties of Entangled Side-Chain Supramolecular Polymer Networks. *Macromolecules* **52**, 5255-5267 (2019).
77. Semenov, A., Charlot, A., Auzély-Velty, R. & Rinaudo, M. Rheological properties of binary associating polymers. *Rheologica Acta* **46**, 541-568 (2007).
78. Wang, S. & Larson, R. G. Multiple relaxation modes in suspensions of colloidal particles bridged by telechelic polymers. *Journal Of Rheology* **62**, 477-490 (2018).
79. Ginzburg, V. V., Chatterjee, T., Nakatani, A. I. & Van Dyk, A. K. Oscillatory and steady shear rheology of model hydrophobically modified ethoxylated urethane-thickened waterborne paints. *Langmuir* **34**, 10993-11002 (2018).
80. Huskens, J., Prins, L. J., Haag, R. & Ravoo, B. J. *Multivalency: Concepts, Research and Applications.* (John Wiley & Sons, 2018).
81. Roy, H. L., Song, J., McKinley, G. H., Holten-Andersen, N. & Lenz, M. Valence can control the non-exponential viscoelastic relaxation of reversible multivalent gels. *arXiv* (2021).
82. Epstein, E. S. *et al.* Modulating noncovalent cross-links with molecular switches. *Journal of the American Chemical Society* **141**, 3597-3604 (2019).
83. Masurel, R. J. *et al.* Role of dynamical heterogeneities on the viscoelastic spectrum of polymers: a stochastic continuum mechanics model. *Macromolecules* **48**, 6690-6702 (2015).
84. Schirmacher, W., Ruocco, G. & Mazzone, V. Theory of heterogeneous viscoelasticity. *Philosophical Magazine* **96**, 620-635 (2016).
85. Ngai, K. *Relaxation and Diffusion in Complex Systems*. (Springer Science & Business Media, 2011).
86. Phillips, J. Stretched exponential relaxation in molecular and electronic glasses. *Reports On Progress In Physics* **59**, 1133 (1996).
87. Berry, G. C. & Plazek, D. J. On the use of stretched-exponential functions for both linear viscoelastic creep and stress relaxation. *Rheologica Acta* **36**, 320-329 (1997).
88. Erk, K. A. & Douglas, J. F. Stretched exponential stress relaxation in a thermally reversible, physically associating block copolymer solution. *MRS Online Proceedings Library Archive* **1418** (2012).
89. Chaudhuri, O. *et al.* Hydrogels with tunable stress relaxation regulate stem cell fate and activity. *Nature Materials* **15**, 326-334 (2016).
90. Bunde, A., Havlin, S., Klafter, J., Graff, G. & Shehter, A. Stretched-exponential relaxation: the role of system size. *Philosophical Magazine B* **77**, 1323-1329 (1998).
91. Gurtovenko, A. A. & Blumen, A. in *Polymer Analysis Polymer Theory* 171-282 (Springer, 2005).
92. Douglas, J. F. & Hubbard, J. B. Semiempirical theory of relaxation: Concentrated polymer solution dynamics. *Macromolecules* **24**, 3163-3177 (1991).
93. Stukalin, E. B., Douglas, J. F. & Freed, K. F. Multistep relaxation in equilibrium polymer solutions: A minimal model of relaxation in "complex" fluids. *The Journal Of Chemical Physics* **129**, 094901 (2008).
94. Curro, J. G. & Pincus, P. A theoretical basis for viscoelastic relaxation of elastomers in the long-time limit. *Macromolecules* **16**, 559-562 (1983).
95. Rubinstein, M. & Obukhov, S. Power-law-like stress relaxation of block copolymers: disentanglement regimes. *Macromolecules* **26**, 1740-1750 (1993).
96. Winter, H. H. & Mours, M. in *Neutron Spin Echo Spectroscopy Viscoelasticity Rheology* 165-234 (Springer, 1997).
97. Tighe, B. P. Relaxations and rheology near jamming. *Physical Review Letters* **107**, 158303 (2011).
98. Yucht, M., Sheinman, M. & Broedersz, C. Dynamical behavior of disordered spring networks. *Soft Matter* **9**, 7000-7006 (2013).
99. Muthukumar, M. Dynamics of polymeric fractals. *The Journal Of Chemical Physics* **83**, 3161-3168 (1985).
100. Muthukumar, M. & Winter, H. H. Fractal dimension of a crosslinking polymer at the gel point. *Macromolecules* **19**, 1284-1285 (1986).
101. Muthukumar, M. Screening effect on viscoelasticity near the gel point. *Macromolecules* **22**, 4656-4658 (1989).
102. Adolf, D. & Martin, J. E. Ultraslow relaxations in networks: evidence for remnant fractal structures. *Macromolecules* **24**, 6721-6724 (1991).
103. Keshavarz, B. *et al.* Time–connectivity superposition and the gel/glass duality of weak colloidal gels. *Proceedings of the National Academy of Sciences* **118**, e2022339118 (2021).
104. Bantawa, M. *et al.* The hidden hierarchical nature of soft particulate gels. *arXiv preprint arXiv:2211.03693* (2022).
105. Head, D. Viscoelastic scaling regimes for marginally rigid fractal spring networks. *Physical Review Letters* **129**, 018001 (2022).
106. Karakoulaki, A. & Head, D. Distinct viscoelastic scaling for isostatic spring networks of the same fractal dimension. *arXiv preprint arXiv:2208.02026* (2022).
107. Varga, Z. & Swan, J. W. Normal modes of weak colloidal gels. *Physical Review E* **97**, 012608 (2018).
108. Dennison, M. & Stark, H. Viscoelastic properties of marginal networks in a solvent. *Physical Review E* **93**, 022605 (2016).
109. Krall, A. & Weitz, D. Internal dynamics and elasticity of fractal colloidal gels. *Physical Review Letters* **80**, 778 (1998).
110. Cho, J. H., Cerbino, R. & Bischofberger, I. Emergence of Multiscale Dynamics in Colloidal Gels. *Physical Review Letters* **124**, 088005 (2020).
111. Cho, J. H. & Bischofberger, I. Two modes of cluster dynamics govern the viscoelasticity of colloidal gels. *Physical Review E* **103**, 032609 (2021).
112. Zaccone, A., Winter, H. H., Siebenbürger, M. & Ballauff, M. Linking self-assembly, rheology, and gel transition in attractive colloids. *Journal Of Rheology* **58**, 1219-1244 (2014).
113. Sollich, P., Lequeux, F., Hébraud, P. & Cates, M. E. Rheology of soft glassy materials. *Physical Review Letters* **78**, 2020 (1997).
114. Fielding, S. M., Sollich, P. & Cates, M. E. Aging and rheology in soft materials. *Journal of Rheology* **44**, 323-369 (2000).
115. Bouchaud, J.-P. Weak ergodicity breaking and aging in disordered systems. *Journal De Physique I* **2**, 1705-1713 (1992).
116. Langer, J. S. Shear-transformation-zone theory of plastic deformation near the glass transition. *Physical Review E* **77**, 021502 (2008).
117. Falk, M. L. & Langer, J. S. Deformation and failure of amorphous, solidlike materials. *Annu. Rev. Condens. Matter Phys.* **2**, 353-373 (2011).
118. Fuereder, I. & Ilg, P. Nonequilibrium thermodynamics of the soft glassy rheology model. *Physical Review E* **88**, 042134 (2013).
119. Bouchbinder, E. & Langer, J. S. Nonequilibrium thermodynamics and glassy rheology. *Soft Matter* **9**, 8786-8791 (2013).
120. Sollich, P. & Cates, M. E. Thermodynamic interpretation of soft glassy rheology models. *Physical Review E* **85**, 031127 (2012).
121. Nicolas, A., Ferrero, E. E., Martens, K. & Barrat, J.-L. Deformation and flow of amorphous solids: Insights from elastoplastic models. *Reviews Of Modern Physics* **90**, 045006 (2018).
122. Ferrero, E. E., Martens, K. & Barrat, J.-L. Relaxation in yield stress systems through elastically interacting activated events. *Physical Review Letters* **113**, 248301 (2014).
123. Shivers, J. L., Arzash, S., Sharma, A. & MacKintosh, F. C. Scaling theory for mechanical critical behavior in fiber networks. *Physical review letters* **122**, 188003 (2019).
124. Rizzi, L., Auer, S. & Head, D. Importance of non-affine viscoelastic response in disordered fibre networks. *Soft Matter* **12**, 4332-4338 (2016).
125. Shivers, J. L., Sharma, A. & MacKintosh, F. C. Nonaffinity controls critical slowing down and rheology near the onset of rigidity. *arXiv preprint arXiv:2203.04891* (2022).
126. Liu, A. J., Ramaswamy, S., Mason, T., Gang, H. & Weitz, D. Anomalous viscous loss in emulsions. *Physical Review Letters* **76**, 3017 (1996).
127. Mulla, Y., MacKintosh, F. & Koenderink, G. H. Origin of slow stress relaxation in the cytoskeleton. *Physical Review Letters* **122**, 218102 (2019).
128. Song, J. *et al.* Microscopic dynamics underlying the stress relaxation of arrested soft materials. *Proceedings of the National Academy of Sciences* **119**, e2201566119 (2022).
129. Shang, B., Guan, P. & Barrat, J.-L. Elastic avalanches reveal marginal behavior in amorphous solids. *Proceedings Of The National Academy Of Sciences* **117**, 86-92 (2020).

130. Cao, P., Short, M. P. & Yip, S. Potential energy landscape activations governing plastic flows in glass rheology. *Proceedings Of The National Academy Of Sciences* **116**, 18790-18797 (2019).
131. Charbonneau, P., Kurchan, J., Parisi, G., Urbani, P. & Zamponi, F. Fractal free energy landscapes in structural glasses. *Nature Communications* **5**, 1-6 (2014).
132. Bouchaud, J.-P. & Pitard, E. Anomalous dynamical light scattering in soft glassy gels. *The European Physical Journal E* **6**, 231-236 (2001).
133. Cipelletti, L. *et al.* Universal non-diffusive slow dynamics in aging soft matter. *Faraday Discussions* **123**, 237-251 (2003).
134. Underhill, P. T. & Graham, M. D. Correlations and fluctuations of stress and velocity in suspensions of swimming microorganisms. *Physics Of Fluids* **23**, 121902 (2011).
135. Hwang, H. J., Riggleman, R. A. & Crocker, J. C. Understanding soft glassy materials using an energy landscape approach. *Nature Materials* **15**, 1031-1036, doi:10.1038/nmat4663 (2016).
136. Lau, A. W., Hoffman, B. D., Davies, A., Crocker, J. C. & Lubensky, T. C. Microrheology, stress fluctuations, and active behavior of living cells. *Physical Review Letters* **91**, 198101 (2003).
137. Baumgaertel, M. & Winter, H. H. Determination of discrete relaxation and retardation time spectra from dynamic mechanical data. *Rheologica Acta* **28**, 511-519 (1989).
138. Baumgaertel, M. & Winter, H. H. Interrelation between continuous and discrete relaxation time spectra. *Journal Of Non-Newtonian Fluid Mechanics* **44**, 15-36 (1992).
139. Jensen, E. A. Determination of discrete relaxation spectra using simulated annealing. *Journal Of Non-Newtonian Fluid Mechanics* **107**, 1-11 (2002).
140. Soo Cho, K. & Woo Park, G. Fixed-point iteration for relaxation spectrum from dynamic mechanical data. *Journal Of Rheology* **57**, 647-678 (2013).
141. Bae, J.-E. & Cho, K. S. Logarithmic method for continuous relaxation spectrum and comparison with previous methods. *Journal Of Rheology* **59**, 1081-1112 (2015).
142. Stadler, F. J. & Bailly, C. A new method for the calculation of continuous relaxation spectra from dynamic-mechanical data. *Rheologica Acta* **48**, 33-49 (2009).
143. Honerkamp, J. & Weese, J. A nonlinear regularization method for the calculation of relaxation spectra. *Rheologica Acta* **32**, 65-73 (1993).
144. Forney, D. C. & Rothman, D. H. Common structure in the heterogeneity of plant-matter decay. *Journal Of The Royal Society Interface* **9**, 2255-2267 (2012).
145. Takeh, A. & Shanbhag, S. A computer program to extract the continuous and discrete relaxation spectra from dynamic viscoelastic measurements. *Applied Rheology* **23** (2013).
146. Emri, I. & Tschoegl, N. Generating line spectra from experimental responses. Part I: Relaxation modulus and creep compliance. *Rheologica Acta* **32**, 311-322 (1993).
147. Anderssen, R. S. & Davies, A. Simple moving-average formulae for the direct recovery of the relaxation spectrum. *Journal of rheology* **45**, 1-27 (2001).
148. Freund, J. B. & Ewoldt, R. H. Quantitative rheological model selection: Good fits versus credible models using Bayesian inference. *Journal of Rheology* **59**, 667-701 (2015).
149. Laun, H. Description of the non-linear shear behaviour of a low density polyethylene melt by means of an experimentally determined strain dependent memory function. *Rheologica Acta* **17**, 1-15 (1978).
150. Winter, H., Baumgaertel, M. & Soskey, P. in *Techniques in rheological measurement* 123-160 (Springer, 1993).
151. Martinetti, L., Soulages, J. M. & Ewoldt, R. H. Continuous relaxation spectra for constitutive models in medium-amplitude oscillatory shear. *Journal Of Rheology* **62**, 1271-1298 (2018).
152. Mainardi, F. On some properties of the Mittag-Leffler function $E_\alpha(-t^\alpha)$, completely monotone for t > 0 with 0<$\alpha$< 1. *Arxiv 1305.0161* (2013).
153. Gorenflo, R., Kilbas, A. A., Mainardi, F. & Rogosin, S. V. *Mittag-Leffler functions, related topics and applications*. Vol. 2 (Springer, 2014).
154. Rogosin, S. The role of the Mittag-Leffler function in fractional modeling. *Mathematics* **3**, 368-381 (2015).
155. Rosa, C. & Capelas de Oliveira, E. Relaxation equations: fractional models. *Journal Of Physical Mathematics* **6** (2015).
156. Katicha, S. W. & Flintsch, G. Fractional viscoelastic models: master curve construction, interconversion, and numerical approximation. *Rheologica Acta* **51**, 675-689 (2012).
157. Garrappa, R., Mainardi, F. & Guido, M. Models of dielectric relaxation based on completely monotone functions. *Fractional Calculus And Applied Analysis* **19**, 1105-1160 (2016).
158. Adolfsson, K., Enelund, M. & Olsson, P. On the fractional order model of viscoelasticity. *Mechanics Of Time-Dependent Materials* **9**, 15-34 (2005).
159. Heymans, N. Hierarchical models for viscoelasticity: dynamic behaviour in the linear range. *Rheologica Acta* **35**, 508-519 (1996).
160. Metzler, R. & Klafter, J. From stretched exponential to inverse power-law: fractional dynamics, Cole–Cole relaxation processes, and beyond. *Journal of Non-Crystalline Solids* **305**, 81-87 (2002).
161. Prabhakar, T. R. A singular integral equation with a generalized Mittag Leffler function in the kernel. *Yokohama Math. J* **19**, 7-15 (1971).
162. Garra, R. & Garrappa, R. The Prabhakar or three parameter Mittag–Leffler function: Theory and application. *Communications In Nonlinear Science And Numerical Simulation* **56**, 314-329 (2018).
163. Metzler, R., Barkai, E. & Klafter, J. Anomalous diffusion and relaxation close to thermal equilibrium: A fractional Fokker-Planck equation approach. *Physical Review Letters* **82**, 3563 (1999).
164. De Oliveira, E. C., Mainardi, F. & Vaz, J. Models based on Mittag-Leffler functions for anomalous relaxation in dielectrics. *The European Physical Journal Special Topics* **193**, 161-171 (2011).
165. Jaishankar, A. & McKinley, G. H. Power-law rheology in the bulk and at the interface: quasi-properties and fractional constitutive equations. *Proceedings Of The Royal Society A* **469**, 20120284 (2013).
166. Schiessel, H. & Blumen, A. Hierarchical analogues to fractional relaxation equations. *Journal Of Physics A: Mathematical And General* **26**, 5057 (1993).
167. Nowick, A. S. & Berry, B. S. Lognormal distribution function for describing anelastic and other relaxation processes I. theory and numerical computations. *IBM Journal Of Research And Development* **5**, 297-311 (1961).
168. Nowick, A. S. & Berry, B. S. Lognormal distribution function for describing anelastic and other relaxation processes II. data analysis and applications. *IBM Journal Of Research And Development* **5**, 312-320 (1961).
169. Wiechert, E. Gesetze der elastischen Nachwirkung für constante Temperatur. *Annalen Der Physik* **286**, 546-570 (1893).
170. Wagner, K. W. Zur theorie der unvollkommenen dielektrika. *Annalen Der Physik* **345**, 817-855 (1913).
171. Schirmacher, W., Ruocco, G. & Mazzone, V. Heterogeneous viscoelasticity: A combined theory of dynamic and elastic heterogeneity. *Physical Review Letters* **115**, 015901 (2015).
172. Flores, R. & Perez, J. Mechanical Spectroscopy of the beta Relaxation in Poly (vinyl chloride). *Macromolecules* **28**, 7171-7179 (1995).
173. Feltham, P. On the representation of rheological results with special reference to creep and relaxation. *British Journal Of Applied Physics* **6**, 26 (1955).
174. Fulchiron, R., Michel, A., Verney, V. & Roustant, J. Correlations between relaxation time spectrum and melt spinning behavior of polypropylene. 1: Calculation of the relaxation spectrum as a log-normal distribution and influence of the molecular parameters. *Polymer Engineering & Science* **35**, 513-517 (1995).
175. Cole, K. S. & Cole, R. H. Dispersion and absorption in dielectrics I. Alternating current characteristics. *The Journal Of Chemical Physics* **9**, 341-351 (1941).
176. Lindsey, C. & Patterson, G. Detailed comparison of the Williams–Watts and Cole–Davidson functions. *The Journal Of Chemical Physics* **73**, 3348-3357 (1980).
177. Shang, B., Rottler, J., Guan, P. & Barrat, J.-L. Local versus global stretched mechanical response in a supercooled liquid near the glass transition. *Physical Review Letters* **122**, 105501 (2019).
178. Gross, B. Dielectric relaxation and the Davidson–Cole distribution function. *Journal Of Applied Physics* **57**, 2331-2333 (1985).
179. Dotson, T. C., Budzien, J., McCoy, J. D. & Adolf, D. B. Cole–Davidson dynamics of simple chain models. *The Journal Of Chemical Physics* **130**, 024903 (2009).
180. Blair, G. S., Veinoglou, B. & Caffyn, J. Limitations of the Newtonian time scale in relation to non-equilibrium rheological states and a theory of quasi-properties. *Proceedings of the Royal Society of London. Series A. Mathematical and Physical Sciences* **189**, 69-87 (1947).
181. Jaishankar, A. *The Linear and Nonlinear Rheology of Multiscale Complex Fluids*, Massachusetts Institute of Technology, (2014).

182. Gross, B. & Fuoss, R. M. Ladder structures for representation of viscoelastic systems. *Journal Of Polymer Science* **19**, 39-50 (1956).
183. Friedrich, C., Schiessel, H. & Blumen, A. in *Rheology Series* Vol. 8 429-466 (Elsevier, 1999).
184. Schiessel, H. & Blumen, A. Mesoscopic pictures of the sol-gel transition: ladder models and fractal networks. *Macromolecules* **28**, 4013-4019 (1995).
185. Bagley, R. L. & Torvik, P. A theoretical basis for the application of fractional calculus to viscoelasticity. *Journal of Rheology* **27**, 201-210 (1983).
186. Abdo, M. S., Shah, K., Wahash, H. A. & Panchal, S. K. On a comprehensive model of the novel coronavirus (COVID-19) under Mittag-Leffler derivative. *Chaos, Solitons & Fractals* **135**, 109867 (2020).
187. Sokolov, I. M., Klafter, J. & Blumen, A. Fractional kinetics. *Physics Today* **55**, 48-54 (2002).
188. Rathinaraj, J. D. J., McKinley, G. H. & Keshavarz, B. Incorporating rheological nonlinearity into fractional calculus descriptions of fractal matter and multi-scale complex fluids. *Fractal and Fractional* **5**, 174 (2021).
189. Winter, H. H. & Chambon, F. Analysis of linear viscoelasticity of a crosslinking polymer at the gel point. *Journal Of Rheology* **30**, 367-382 (1986).
190. Celli, J. P. *et al.* Rheology of gastric mucin exhibits a pH-dependent sol− gel transition. *Biomacromolecules* **8**, 1580-1586 (2007).
191. Gross, B. On creep and relaxation. *Journal Of Applied Physics* **18**, 212-221 (1947).
192. Legrand, G., Manneville, S., McKinley, G. H. & Divoux, T. Dual origin of viscoelasticity in polymer-carbon black hydrogels: a rheometry and electrical spectroscopy study. *arXiv preprint arXiv:2210.03606* (2022).
193. Aime, S., Cipelletti, L. & Ramos, L. Power law viscoelasticity of a fractal colloidal gel. *Journal Of Rheology* **62**, 1429-1441 (2018).
194. Sadman, K. *et al.* Influence of hydrophobicity on polyelectrolyte complexation. *Macromolecules* **50**, 9417-9426 (2017).
195. Rathinaraj, J. D. J., Hendricks, J., McKinley, G. H. & Clasen, C. OrthoChirp: A fast spectro-mechanical probe for monitoring transient microstructural evolution of complex fluids during shear. *Journal of Non-Newtonian Fluid Mechanics* **301**, 104744 (2022).
196. Rathinaraj, J. D. J., Keshavarz, B. & McKinley, G. H. Why the Cox–Merz rule and Gleissle mirror relation work: A quantitative analysis using the Wagner integral framework with a fractional Maxwell kernel. *Physics of Fluids* **34**, 033106 (2022).
197. Faber, T., Jaishankar, A. & McKinley, G. Describing the firmness, springiness and rubberiness of food gels using fractional calculus. Part I: Theoretical framework. *Food Hydrocolloids* **62**, 311-324 (2017).
198. Faber, T., Jaishankar, A. & McKinley, G. Describing the firmness, springiness and rubberiness of food gels using fractional calculus. Part II: Measurements on semi-hard cheese. *Food Hydrocolloids* **62**, 325-339 (2017).
199. Song, J. *et al.* Programmable Anisotropy and Percolation in Supramolecular Patchy Particle Gels. *ACS Nano* **14**, 17018-17027 (2020).
200. Singh, P. K., Soulages, J. M. & Ewoldt, R. H. On fitting data for parameter estimates: residual weighting and data representation. *Rheologica Acta* **58**, 341-359 (2019).
201. Singh, P. K. *Rheological Inferences with Uncertainty Quantification*, University of Illinois at Urbana-Champaign, (2019).

**Table 1.** Summary of commonly used fractional mechanical models, their constitutive relations, and analytical responses to common rheological deformations

| Model | Characteristic Modulus $G_c$ Relaxation Time $\tau_c$ | Constitutive Relation | Step Strain Response $\gamma(t) = \gamma_0 \mathcal{H}(t)$ | Oscillatory Strain Response $\gamma(t) = \gamma_0 \sin(\omega t)$ | Step Stress Response $\sigma(t) = \sigma_0 \mathcal{H}(t)$ | Continuous Relaxation Spectrum |
|---|---|---|---|---|---|---|
| **Maxwell Model** 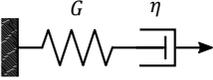 | $\tau_c = \dfrac{\eta}{G}$ $G_c = G$ | $\sigma(t) + \dfrac{\eta}{G}\dfrac{d\sigma(t)}{dt} = \eta \dfrac{d\gamma(t)}{dt}$ | $G(t) = G_c \exp(-t/\tau_c)$ | $\dfrac{G'(\omega)}{G_c} = \dfrac{(\omega\tau_c)^2}{1+(\omega\tau_c)^2}$ $\dfrac{G''(\omega)}{G_c} = \dfrac{\omega\tau_c}{1+(\omega\tau_c)^2}$ | $J(t) = \dfrac{t}{\eta} + \dfrac{1}{G}$ | $H(\tau) = G_c \delta(\tau - \tau_0)$ |
| **Spring-Pot** 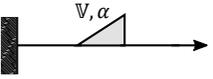 | $G_c \tau_c^\alpha = \mathbb{V}$ | $\sigma(t) = \mathbb{V}\dfrac{d^\alpha \gamma(t)}{dt^\alpha}$ | $G(t) = \mathbb{V}\dfrac{t^{-\alpha}}{\Gamma(1-\alpha)}$ | $G'(\omega) = \mathbb{V}\omega^\alpha \cos(\pi\alpha/2)$ $G''(\omega) = \mathbb{V}\omega^\alpha \sin(\pi\alpha/2)$ | $J(t) = \dfrac{1}{\mathbb{V}}\dfrac{t^\alpha}{\Gamma(1+\alpha)}$ | $H(\tau) = \mathbb{V}\dfrac{\sin\pi\alpha}{\pi\tau^\alpha}$ |
| **Fractional Maxwell Gel** 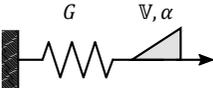 | $\tau_c = \left(\dfrac{\mathbb{V}}{G}\right)^{\frac{1}{\alpha}}$ $G_c = \mathbb{V}\tau_c^{-\alpha} \equiv G$ | $\sigma(t) + \dfrac{\mathbb{V}}{G}\dfrac{d^\alpha \sigma(t)}{dt^\alpha} = \mathbb{V}\dfrac{d^\alpha \gamma(t)}{dt^\alpha}$ | $G(t) = G_c E_{a,b}(z)$ $a = \alpha$ $b = 1$ $z = (-(t/\tau_c)^\alpha)$ | $\dfrac{G'(\omega)}{G_c} = \dfrac{(\omega\tau_c)^{2\alpha} + (\omega\tau_c)^\alpha \cos(\pi\alpha/2)}{1+(\omega\tau_c)^{2\alpha} + 2(\omega\tau_c)^\alpha \cos(\pi\alpha/2)}$ $\dfrac{G''(\omega)}{G_c} = \dfrac{(\omega\tau_c)^\alpha \sin(\pi\alpha/2)}{1+(\omega\tau_c)^{2\alpha} + 2(\omega\tau_c)^\alpha \cos(\pi\alpha/2)}$ | $J(t) = \dfrac{1}{\mathbb{V}}\dfrac{t^\alpha}{\Gamma(1+\alpha)} + \dfrac{1}{G}$ | $H(\tau) = \dfrac{G_c}{\pi}\dfrac{(\tau/\tau_c)^\alpha \sin(\pi\alpha)}{(\tau/\tau_c)^{2\alpha} + 2(\tau/\tau_c)^\alpha \cos(\pi\alpha) + 1}$ |
| **Fractional Maxwell Liquid** 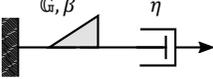 | $\tau_c = \left(\dfrac{\eta}{\mathbb{G}}\right)^{\frac{1}{1-\beta}}$ $G_c = \eta\tau_c^{-1}$ | $\sigma(t) + \dfrac{\eta}{\mathbb{G}}\dfrac{d^{1-\beta}\sigma(t)}{dt^{1-\beta}} = \eta \dfrac{d\gamma(t)}{dt}$ | $G(t) = G_c(t/\tau_c)^{-\beta} E_{a,b}(z)$ $a = 1-\beta$ $b = 1-\beta$ $z = (-(t/\tau_c)^{1-\beta})$ | $\dfrac{G'(\omega)}{G_c} = \dfrac{(\omega\tau_c)^{2-\beta}\cos(\pi\beta/2)}{1+(\omega\tau_c)^{2(1-\beta)} + 2(\omega\tau_c)^{1-\beta}\cos(\pi(1-\beta)/2)}$ $\dfrac{G''(\omega)}{G_c} = \dfrac{(\omega\tau_c) + (\omega\tau_c)^{2-\beta}\sin(\pi\beta/2)}{1+(\omega\tau_c)^{2(1-\beta)} + 2(\omega\tau_c)^{1-\beta}\cos(\pi(1-\beta)/2)}$ | $J(t) = \dfrac{t}{\eta} + \dfrac{1}{\mathbb{G}}\dfrac{t^\beta}{\Gamma(1+\beta)}$ | $H(\tau) = \dfrac{G_c}{\pi}\dfrac{(\tau/\tau_c)^{-1}\sin(\pi\beta)}{(\tau/\tau_c)^{\beta-1}+(\tau/\tau_c)^{1-\beta}+2\cos(\pi(1-\beta))}$ |
| **Fractional Maxwell Model** 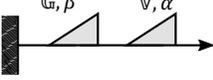 | $\tau_c = \left(\dfrac{\mathbb{V}}{\mathbb{G}}\right)^{\frac{1}{\alpha-\beta}}$ $G_c = \mathbb{V}\tau_c^{-\alpha} \equiv \left(\dfrac{\mathbb{G}^\alpha}{\mathbb{V}^\beta}\right)^{1/(\alpha-\beta)}$ | $\sigma(t) + \dfrac{\mathbb{V}}{\mathbb{G}}\dfrac{d^{\alpha-\beta}\sigma(t)}{dt^{\alpha-\beta}} = \mathbb{V}\dfrac{d^\alpha \gamma(t)}{dt^\alpha}$ | $G(t) = G_c(t/\tau_c)^{-\beta} E_{a,b}(z)$ $a = \alpha-\beta$ $b = 1-\beta$ $z = (-(t/\tau_c)^{\alpha-\beta})$ | $\dfrac{G'(\omega)}{G_c} = \dfrac{(\omega\tau_c)^\alpha \cos(\pi\alpha/2) + (\omega\tau_c)^{2\alpha-\beta}\cos(\pi\beta/2)}{1+(\omega\tau_c)^{2(\alpha-\beta)} + 2(\omega\tau_c)^{\alpha-\beta}\cos(\pi(\alpha-\beta)/2)}$ $\dfrac{G''(\omega)}{G_c} = \dfrac{(\omega\tau_c)^\alpha \sin(\pi\alpha/2) + (\omega\tau_c)^{2\alpha-\beta}\sin(\pi\beta/2)}{1+(\omega\tau_c)^{2(\alpha-\beta)} + 2(\omega\tau_c)^{\alpha-\beta}\cos(\pi(\alpha-\beta)/2)}$ | $J(t) = \dfrac{1}{\mathbb{V}}\dfrac{t^\alpha}{\Gamma(1+\alpha)} + \dfrac{1}{\mathbb{G}}\dfrac{t^\beta}{\Gamma(1+\beta)}$ | $H(\tau) = \dfrac{G_c}{\pi}\dfrac{(\tau/\tau_c)^{-\beta}\sin(\pi\alpha) + (\tau/\tau_c)^{-\alpha}\sin(\pi\beta)}{(\tau/\tau_c)^{\beta-\alpha}+(\tau/\tau_c)^{\alpha-\beta}+2\cos(\pi(\alpha-\beta))}$ |
| **Fractional Kelvin-Voigt Model** 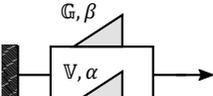 | $\tau_c = \left(\dfrac{\mathbb{V}}{\mathbb{G}}\right)^{\frac{1}{\alpha-\beta}}$ $G_c = \mathbb{V}\tau_c^{-\alpha} \equiv \left(\dfrac{\mathbb{G}^\alpha}{\mathbb{V}^\beta}\right)^{1/(\alpha-\beta)}$ | $\sigma(t) = \mathbb{V}\dfrac{d^\alpha \gamma(t)}{dt^\alpha} + \mathbb{G}\dfrac{d^\beta \gamma(t)}{dt^\beta}$ | $G(t) = \mathbb{V}\dfrac{t^{-\alpha}}{\Gamma(1-\alpha)} + \mathbb{G}\dfrac{t^{-\beta}}{\Gamma(1-\beta)}$ | $G'(\omega) = \mathbb{V}\omega^\alpha \cos(\pi\alpha/2) + \mathbb{G}\omega^\beta \cos(\pi\beta/2)$ $G''(\omega) = \mathbb{V}\omega^\alpha \sin(\pi\alpha/2) + \mathbb{G}\omega^\beta \sin(\pi\beta/2)$ | $J(t) = \dfrac{t^\alpha}{\mathbb{V}}E_{a,b}(z)$ $a = \alpha-\beta$ $b = 1+\alpha$ $z = (-(t/\tau_c)^{\alpha-\beta})$ | $H(\tau) = \mathbb{V}\dfrac{\sin\pi\alpha}{\pi\tau^\alpha} + \mathbb{G}\dfrac{\sin\pi\beta}{\pi\tau^\beta}$ |

**Notations -** $G(t)$: Stress relaxation modulus;   $G'(\omega)$, $G''(\omega)$: Storage and loss modulus;   $J(t)$: Creep compliance;   $H(\tau)$: Relaxation spectrum

**Mathematical Functions -** Delta function $\int_{0_-}^{0^+} \delta(x)\,dx = 1$;   Gamma function $\Gamma(x) = (x-1)!$;   Mittag-Leffler function $E_{a,b}(z) = \sum_{n=0}^{\infty}\left[\dfrac{z^n}{\Gamma(an+b)}\right]$

**Supplementary Information**

Table S1. Fitting parameters for the data in the manuscript

| Data | Function | Fitting parameters ([ ]: Units) | Parameter values |
|---|---|---|---|
| Stress relaxation of nanoparticle-polymer gel (Fig. 4A) | Stretched exponential function (Equation 9) | $\{G_0 [Pa], \tau_c\ [s], \alpha\}$ | $\{1.05, 3.91 \times 10^3, 0.31\}$ |
| Dynamic modulus of metal-coordinating polymer networks (Fig. 4C) | Log-normal $H(\tau)$ function (Equation 17) | $\{G_0 [Pa], \tau_c\ [s], \sigma_w\}$ | $\{3.54 \times 10^4, 8.84 \times 10^{-5}, 4.47\}$ |
| Dynamic modulus of critical mucin gels (Fig. 6A) | Spring-pot (Table 1) | $\{\mathbb{V}\ [Pa\ s^\alpha], \alpha\}$ | $\{0.41, 0.39\}$ |
| Dynamic modulus of silica colloidal gels (Fig. 6B) | Fractional Maxwell gel (Table 1) | $\{G\ [Pa], \mathbb{V}\ [Pa\ s^\alpha], \alpha\}$ | $\{2.76 \times 10^3, 8.84 \times 10^4, 0.36\}$ |
| Dynamic modulus of polyelectrolyte complexes (Fig. 6C) | Fractional Maxwell liquid (Table 1) | $\{\eta\ [Pa\ s], \mathbb{G}\ [Pa\ s^\beta], \beta\}$ | $\{25.45, 744.10, 0.35\}$ |
| Stress relaxation of fish muscle tissue (Fig. 6D) | Fractional Maxwell model (Table 1) | $\{\mathbb{V}\ [Pa\ s^\alpha], \mathbb{G}\ [Pa\ s^\beta], \alpha, \beta\}$ | $\{8.75 \times 10^5, 2.21 \times 10^4, 0.68, 0.11\}$ |

**Demonstration of MATLAB codes for fitting rheological data**

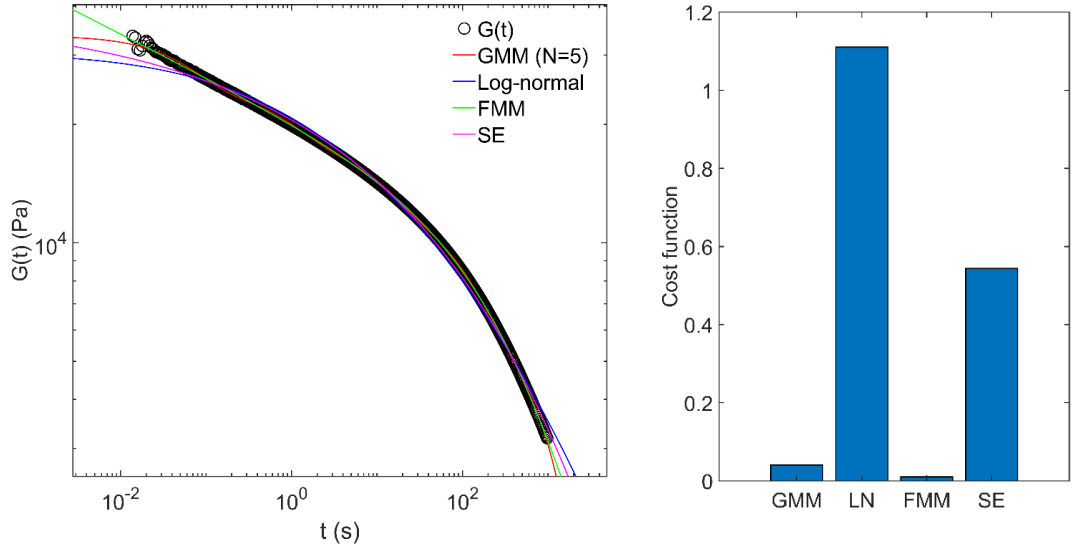

**Figure S1. Fit results for $G(t)$ data from measurements on yellowfish tuna myotome.** (Left) Fit results to the data. (Right) Cost function computed from the weighted residual sum of squares. See attached MATLAB demo for details on the fitting procedures.

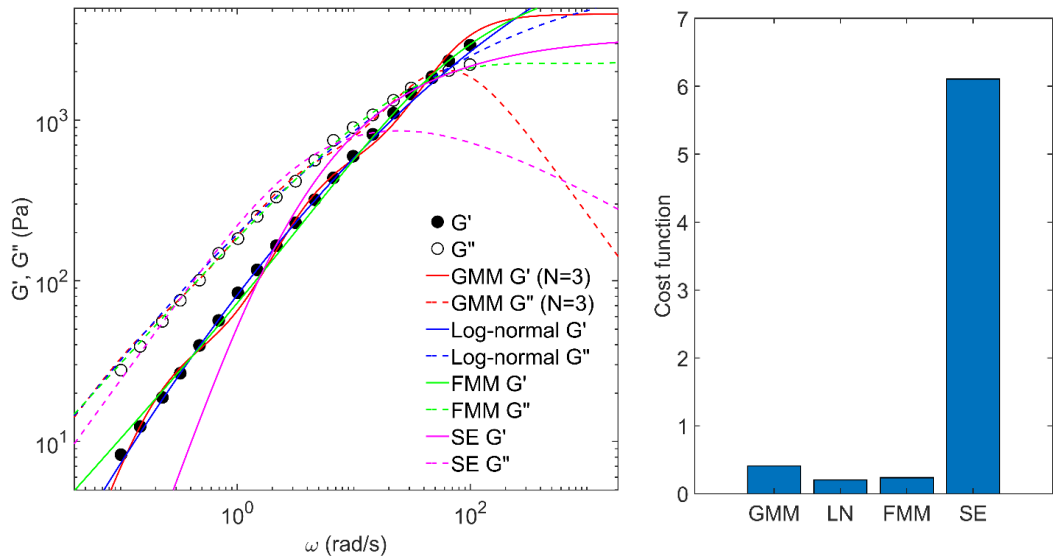

**Figure S2. Fit results for the $G'(\omega)$ and $G''(\omega)$ data of Epstein et al.** (Left) Fit results to the data of Epstein et al., (Right) Cost function computed from the weighted residual sum of squares. See attached MATLAB demo for details on the fitting procedures.